%% file: ms.tex
\pdfoutput=1
\RequirePackage{fixltx2e}[2005/12/01] 
\documentclass[useAMS,usenatbib]{mn2e}

\usepackage[pdftex]{graphicx}
\usepackage{amssymb}
\usepackage{amsmath}
\usepackage{amstext}
\usepackage{multirow} 
\usepackage{rotating}

\usepackage{xcolor}

\usepackage[varg]{txfonts}
\usepackage{microtype}

\usepackage[autolanguage]{numprint}
\npfourdigitnosep
\newcommand\num{\numprint}
\newcommand\SI[2]{\numprint[#2]{#1}}

\newcommand\hbeta{\ensuremath{\mathrm{H}\beta}}
\newcommand\halpha{\ensuremath{\mathrm{H}\alpha}}
\newcommand\elec{\ensuremath{_{\mathrm{e}}}}
\newcommand\crit{\ensuremath{^{\mathrm{crit}}}}
\newcommand\proplyd{\ensuremath{^{\mathrm{\,p}}}}
\newcommand\pe{\proplyd}        
\newcommand\total{\ensuremath{^{\mathrm{\,t}}}}
\newcommand\bg{\ensuremath{^{\mathrm{\,b}}}}

\newcommand\los{\ensuremath{^{\mathrm{los}}}}
\newcommand\FWHM{\ensuremath{\mathrm{FWHM}}}
\newcommand\Ion[2]{\ensuremath{\mathrm{#1\,\textsc{#2}}}}
\newcounter{ionstage}
\newcommand{\ionW}[2]{
  \setcounter{ionstage}{#2}%
  \Ion{#1}{\roman{ionstage}}}

\newcommand\ioni[2]{\ensuremath{\mathrm{#1}^{#2}}}           

\newcommand{\adfo}{\ensuremath{\mathrm{ADF}(\ioni{O}{2+})}}

\newcommand{\te}{\ensuremath{T\elec}}
\newcommand{\nel}{\ensuremath{n\elec}}

\newcommand{\chb}{\ensuremath{c(\hbeta)}}
\newcommand{\tf}{\ensuremath{t^{2}}}
\let\ion=\Ion                                  
\newcommand\wav[1]{\ensuremath{\lambda #1}}
\newcommand\wavs[2]{\ensuremath{\lambda\lambda #1,#2}}

\newcommand\thC{\(\theta^1\,\)Ori~C}
\providecommand\arcdeg{\degr} 

\newcommand\unit[1]{\ensuremath{\mathrm{#1}}}
\newcommand\K{\unit{K}}
\newcommand\cmc{\unit{cm^{-3}}}
\newcommand\kms{\unit{km\ s^{-1}}}
\newcommand\pc{\unit{pc}}

\newcommand{\hii}{\ion{H}{ii}}

\newcommand\HMSf[4]{\ensuremath{#1^{\mathrm{h}}\ #2^{\mathrm{m}}\ #3\fs #4}}
\newcommand\DMS[3]{\ensuremath{#1\degr\ #2\arcmin\ #3\arcsec}}

\renewcommand\fs{\ensuremath{\rlap{\(\mathrm{^s}\)}.}} 
\newcommand\arcseckern{\kern-1pt}                      
\newlength\arcsecwidth
\renewcommand\farcs{\ensuremath{\rlap{\makebox[\arcsecwidth][c]{.}}\arcsec\arcseckern}}

\newcommand\fnm[1]{\textsuperscript{#1}} 
\newcommand\tfnm[1]{\fnm{\textit{#1}}} 

\newcommand\mykern{\kern-0.2em}
\def\coords(#1:#2,#3:#4){\ensuremath{(#1 \mykern:\mykern #2, #3 \mykern:\mykern #4)}\relax}

\title[Integral field spectroscopy of proplyds]{%
  Ionized gas diagnostics from protoplanetary discs in the Orion Nebula 
  and the abundance discrepancy problem%
  \thanks{Based on observations collected at the Centro Astron\'omico 
    Hispano Alem\'an (CAHA) at Calar Alto, operated jointly by the 
    Max-Planck Institut f\"ur Astronomie and the Instituto de 
    Astrof{\'{\i}}sica de Andaluc{\'{\i}}a (CSIC).}
}

\author[A. Mesa-Delgado et al.]{%
  A. Mesa-Delgado\thanks{E-mail: amesad@astro.puc.cl}\fnm{1,2,3,4}, 
  M. N\'u\~nez-D{\'{i}}az\fnm{3,4}, C. Esteban\fnm{3,4}, J. Garc\'ia-Rojas\fnm{3,4}, 
  \newauthor 
  N. Flores-Fajardo\fnm{5},  
  L. L\'opez-Mart{\'{\i}}n\fnm{3,4}, Y.~G. Tsamis\fnm{6} and W.~J. Henney\fnm{5}\\  
  \fnm{1}Departamento de Astronom\'ia y Astrof\'isica, 
  Facultad de F\'isica, Pontificia Universidad Cat\'olica de 
  Chile, Av.~Vicu\~na Mackenna 4860,\\ 782-0436 Macul, Santiago, Chile\\
  \fnm{2}Institute for Astronomy, 2680 Woodlawn Drive, Honolulu, HI 96822, USA\\       
  \fnm{3}Instituto de Astrof\'\i sica de Canarias, E-38200 La Laguna, Tenerife, Spain \\
  \fnm{4}Departamento de Astrof\'\i sica, Universidad de La Laguna, E-38205 La Laguna, 
  Tenerife, Spain \\
  \fnm{5}Centro de Radioastronom\'{\i}a y Astrof\'{\i}sica, 
  Universidad Nacional Aut\'onoma de M\'exico,
  Campus Morelia, Apartado Postal 3-72, \\ 58090 Morelia, Michoac\'an, M\'exico\\
  \fnm{6}European Southern Observatory, Karl-Schwarzschild-Str. 2, D-85748 Garching bei M\"unchen, 
  Germany\\
}

\begin{document}

\date{Accepted 2012 May 2.  Received 2012 May 2; in original form 2012 March 14}

\maketitle
\label{firstpage}

\begin{abstract}
  We present results from integral field spectroscopy of a field located near the Trapezium Cluster using the Potsdam Multi-Aperture Spectrograph (PMAS). 
  The observed field contains a variety of morphological structures: five externally ionized protoplanetary discs (also known as proplyds), the high-velocity jet HH~514 and a bowshock. 
  Spatial distribution maps are obtained for different emission line fluxes, the \chb\ extinction coefficient, electron densities and temperatures, ionic abundances of different ions from collisionally excited lines (CELs), \ioni{C}{2+} and \ioni{O}{2+} abundances from recombination lines (RLs) and the abundance discrepancy factor of \ioni{O}{2+}, \adfo. 
  We distinguish the three most prominent proplyds (177-341, 170-337 and 170-334) and analyse their impact on the spatial distributions of the above mentioned quantities.
  We find that collisional de-excitation has a major influence on the line fluxes in the proplyds.
  If this is not properly accounted for then physical conditions deduced from commonly used line ratios will be in error, leading to unreliable chemical abundances for these objects.
  We obtain the intrinsic emission of the proplyds 177-341, 170-337 and 170-334 by a direct subtraction of the background emission, though the last two present some background contamination due to their small sizes.
  A detailed analysis of 177-341 spectra making use of suitable density diagnostics reveals the presence of high-density gas (\(3.8\times10^5~\cmc\)) in contrast to the typical values observed in the background gas of the nebula (\(3800~\cmc\)).
  We also explore how the background subtraction could be affected by the possible opacity of the proplyd and its effect on the derivation of physical conditions and chemical abundances of the proplyd 177-341. 
  We construct a physical model for the proplyd 177-341 finding a good agreement between the predicted and observed line ratios. 
  Finally, we find that the use of reliable physical conditions returns an \adfo\ about zero for the intrinsic spectra of 177-341, while the background emission presents the typical \adfo\ observed in the Orion Nebula (\(0.16 \pm 0.11\)~dex). 
  We conclude that the presence of high-density ionized gas is severely affecting the abundances determined from CELs and, therefore, those from RLs should be considered as a better approximation to the true abundances.
\end{abstract}


\section{Introduction} \label{intro} 
Circumstellar discs are very common around young stars and it is well established that they are an integral part of the star-formation and planet-formation processes \citep{williamscieza11}. 
Rotationally supported and geometrically thin discs of dense molecular gas and dust persist throughout the classical T~Tauri stage of low-mass pre-main sequence stellar evolution \citep{beckwithetal90}, with a duration of several million years. 
When such discs are found in or near an \hii\ region their detection, structure and evolution are strongly influenced by the ultraviolet radiation from hot massive stars that dissociates and ionizes the local interstellar medium. 
The currently accepted model \citep{sutherland97, henneyarthur98,  johnstoneetal98, richlingyorke98, storzerhollenbach99, nguyenetal02, vasconcelosetal11} is that far ultraviolet photons heat the surface of the disc, forming a warm outflowing envelope of neutral gas that becomes photoionized after expanding to a few times the outer radius of the disc.  A schematic illustration of this photoevaporating flow model is shown in Fig.~5 of \citet{henneyodell99}. 
This special class of protoplanetary disc is today known as \textsc{proplyd},
a term that was coined by \cite{odelletal93} to describe the tear-drop shaped objects observed in the first imaging studies of Orion Nebula using the Hubble Space Telescope (\textit{HST}). 
Its proximity \citep[\(436 \pm 20~\pc\);][]{odellhenney08} and high surface brightness make the Orion Nebula an important laboratory for the detailed study of phenomena associated with star-formation and the early stages of stellar evolution. 
Indeed, the majority of currently known proplyds \citep[\(\simeq 200\) objects,][]{riccietal08} have been discovered in the Orion Nebula, although an increasing number are now being detected in other \hii\ regions \citep{stapelfeldtetal97, stecklumetal98, brandneretal00, smithetal03, yusefzadehetal05, demarcoetal06, wrightetal12}.
In the central part of the Orion Nebula, the Huygens region, the main features of the proplyds are the brightest zone of the ionization front, known as \textit{cusp}, and the dusty tails of dense gas facing away from \thC{}, the principal ionizing star of the Orion Nebula.
However, what we see depends strongly on the orientation and localization of the proplyds, appearing for example as dark silhouettes against the bright nebular background when found at a distance from \thC{} or within the foreground Orion Veil \citep{mccaughreanodell96, odellwong96}.
 
The first observations of the Orion proplyds were made by \cite{laquesvidal79} who detected six of them around \thC{} as unresolved nebular condensations (given LV designations) with strong emission in \halpha{}, \hbeta{} and [\ion{O}{iii}] \wav{5007}.
From the \halpha{} surface brightness of the condensations, \cite{laquesvidal79} determined electron densities ranging from \(1.5\times10^5\) to \(4.1\times10^5~\cmc\), although these were only lower limits since the objects were not fully resolved.
The correct interpretation of the objects as externally ionized circumstellar discs was initially proposed on the basis of radio interferometric mapping \citep{churchwelletal87} and ground-based optical spectroscopy \citep{meaburn88}, and this was well confirmed by subsequent high-resolution imaging with the \textit{HST} \citep{odelletal93, odellwen94} and ground-based adaptive optics systems \citep{mcculloughetal95}. 
The distribution, evolution and structure of proplyds and their embedded discs have been studied via imaging at  optical, infrared and radio wavelengths \citep[e.g.][]{churchwelletal87, chenetal98, ballyetal98, ballyetal00, ladaetal00, robbertoetal02, smithetal05, eisnercarpenter06,eisneretal08}.
Many spectroscopic studies have also been carried out, with the earliest being devoted to the kinematical properties of the LV knots through  analyzing the [\ion{O}{iii}] \wav{5007} emission profile with high-resolution echelle spectroscopy \citep{meaburn88, meaburnetal93, masseymeaburn93, masseymeaburn95} and comparing these results with the first theoretical models \citep{henneyetal97}.
\cite{ballyetal98} carried out intermediate resolution spectroscopy with FOS at \textit{HST} in the ultraviolet range 1150-3300 \AA\ for proplyds 159-350 and 158-327, detecting \ion{C}{iii}], \ion{C}{ii}], \ion{O}{ii}] and \ion{Mg}{ii} emission.
\cite{henneyodell99} presented a detailed analysis of four proplyds (170-337, 177-341, 182-413, and 244-440) performing high-resolution echelle spectroscopy with the 10m Keck Telescope.
This study analysed the spatial profiles of several emission lines, \hbeta{}, [\ion{O}{iii}] \wav{4959}, \ion{He}{i} \wav{5676}, [\ion{O}{i}] \wav{6300}, [\ion{S}{iii}] \wav{6312}, \halpha{}, [\ion{N}{ii}] \wav{6583} and [\ion{S}{ii}] \wav{6731}, as well as the comparison with theoretical models providing direct measurements of different proplyd properties such as size, ionized density and outflow velocity.
Particularly, \cite{henneyodell99} estimated peak densities in the cusps of about \(10^6~\cmc\).
 
The largest number of detailed spectroscopic studies to date have concentrated on the proplyd LV2 (167-317), which is the most luminous of the inner group of proplyds (separation from \thC{} is \(< 10\arcsec\)). 
\citet{henneyetal02} carried out high-resolution (\(0.05\arcsec \times 2~\kms\)) NUV long-slit spectroscopy with the STIS spectrograph, determining a peak electron density of \(\simeq 3 \times 10^6~\cmc\) in the cusp of this proplyd from the \([\ionW{C}{3}]\ \wav{1907} / \ionW{C}{3}]\ \wav{1909}\) line ratio.
\citet{vasconcelosetal05} performed ground-based integral field spectroscopy of LV2 with GMOS on the 8m Gemini Telescope, detecting multiple forbidden and permitted emission lines in the spectral range from 5500 to 7600~\AA{} and analysing the high-velocity emission associated with the jet \citep[HH~514;][]{meaburn88} that emerges from this proplyd.
More recently, \citet{tsamisetal11} and \citet{tsamiswalsh11} have obtained the deepest spectroscopic observations of LV2 and its redshifted jet with the integral field spectrograph FLAMES on the 8m VLT in the spectral range from 3000 to 7200~\AA.
The FLAMES data were complemented with analysis of the ultraviolet and optical spectra observed with FOS at \textit{HST}.
The high quality of the dataset allowed these authors to perform the first physical and chemical analysis of a proplyd, simultaneously determining electron densities and temperatures from several indicators as well as ionic and total abundances of several elements in the cusp, tail and jet of LV2.
The authors found a high density contrast between the cusp (\(\sim 7.9\times10^5~\cmc\)), the tail (\(\sim \num{38000}~\cmc\)) and the adjacent nebular background (\(\sim 4700~\cmc\)), while the temperatures remain similar for the different structures.
They also found that the gaseous heavy-element abundances are generally higher in the proplyd compared to the nebular background of the Orion Nebula.
The positive correlation found between the host star metallicity and the presence of giant planetary companions \citep[e.g.][and references therein]{nevesetal09} has aroused a great interest in the question of the chemical composition of planet-formation circumstellar envelopes.
It is therefore important to check if the observed abundance difference between LV2 and the Orion Nebula is a general property of the proplyds.
  \begin{figure*}
   \centering
   \includegraphics[scale=2.6]{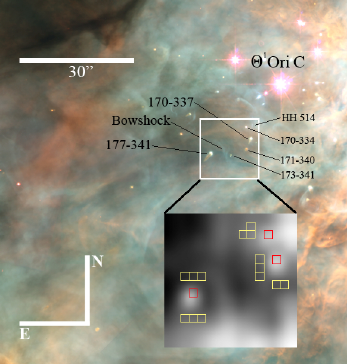} 
   \caption{Section of the central part of the Orion Nebula, a combination of the \textit{HST}/WFPC2 images taken with different filters \citep{odellwong96}. The PMAS field of view of \(16 \arcsec \times 16\arcsec\) is shown by the white square. The inset zooms in the PMAS image in \halpha{} were rebinned to \(160 \times 160\) pixels and smoothed using a Gaussian filter. The red boxes represent the extracted spaxels to analyse the emission from the proplyds 177-341, 170-337 and 170-334, while the yellow boxes correspond to their adopted nebular backgrounds.}
   \label{f1}
  \end{figure*}

The chemical analysis carried out by \cite{tsamisetal11} shows that a detailed study of proplyd spectra is a way to investigate the so-called abundance discrepancy (AD) problem.
This is currently the most controversial problem in nebular astrophysics \citep[see][]{liu02, esteban02, tsamisetal03, peimbertapeimbert06, liu06, stasinskaetal07, garciarojasesteban07, rodriguezgarciarojas10, simondiazstasinska11}.
The AD is the disagreement between the abundances of the same ion derived from collisionally excited lines (CELs) and recombination lines (RLs) of heavy elements.
For instance, \cite{garciarojasesteban07} found that, 
for a large sample of Galactic and extragalactic \hii\ regions, 
the \ioni{O}{2+}/\ioni{H}{+} ratio calculated from RLs is always between 0.1 and 0.3~dex higher than the value obtained from CELs for the same region.
This difference is known as the abundance discrepancy factor: \adfo.  
For the Orion Nebula, numerous studies have shown that \adfo{} is \(\sim 0.15\)~dex for the background nebular gas \citep[e.g.][]{peimbertetal93, estebanetal98, estebanetal04, mesadelgadoetal08}.
An initial attempt to investigate the possible role of the proplyds in the AD problem \citep{mesadelgadoetal08} did not give a clear answer due to the presence of absorption features around the \ion{O}{ii} RLs in the proplyd spectra, the absence of proper high-density indicators, and the impossibility of isolating the proplyd emission using long-slit spectroscopy \citep[see][]{henneyodell99}.
A striking result found by \cite{tsamisetal11} is that the \adfo\ is virtually zero in the cusp due to the high density of the proplyd, which produces a clear enhancement of the \ioni{O}{2+} abundances from CELs with respect to the nebular background, while those from RLs are basically similar in both cases.  
 
The main aim of the present paper is to study several proplyds in the Orion Nebula
using integral field spectroscopy in the optical range at spatial scales of about 1\arcsec\ and explore their impact on the AD problem. 
At the same time, we derive the physical conditions in the proplyds from empirical diagnostics of temperature, density, and ionization.  
In \S\ref{obsred} we describe the observations and the data reduction procedure.
In \S\ref{medpmas} the emission line measurements and the reddening correction as well as some representative maps of those quantities are presented.
In \S\ref{results} we analyse the spatial distributions of the excitation ratios, physical conditions and chemical abundances. 
We also attempt to separate the intrinsic emission of the proplyds from that of the background nebula. 
The effect of uncertainties in the dust opacity of the proplyd on the background subtraction procedure is discussed in \S\ref{dust}. 
In \S\ref{model} we present preliminary results of a physical model for the proplyd 177-341 and discuss how these complement the more empirical analysis. 
Finally, in \S\ref{conclu} our main conclusions are summarized. 
  
\section{Observations, data reduction and extraction of proplyd spectra} \label{obsred}  
\subsection{Observations and data reduction}
Integral field spectroscopy was carried out on 2008 December 19 with the Potsdam Multi-Aperture Spectrometer \citep[PMAS,][]{rothetal2005} on the 3.5m telescope at the Calar Alto Observatory (Almer\'ia, Spain).
The standard lens array integral field unit (IFU) of \(16\arcsec\times 16 \arcsec\) field of view (FoV) with a spaxel size of \(1\arcsec \times 1\arcsec\) was used.
The V600 grating was used at two rotator angles to cover the blue part of the spectrum from 3500 to 5100~\AA{} (angle of \(-72\arcdeg\)) and the red part of the spectrum from 5700 to 7200~\AA{} (angle of \(-68\arcdeg\)).

We achieved an effective spectral resolution \(\FWHM \sim 3.6~\AA\) in both spectral ranges.
The total integration times for the blue and red spectra were \(3 \times 400\)~s and 300~s, respectively.
Additional short 10 second exposures were also taken for each range, avoiding the saturation of the brightest emission lines.
Different calibration images were also taken: HgNe arc lamps for the wavelength calibration and continuum lamps needed to extract the 256 individual spectra on the CCD.
The spectrophotometric standard stars Feige~100, Feige~34, Feige~110 and G~191-B2B \citep{oke90} were observed to perform the flux calibration.
The error of the absolute flux calibration was about 5\%.
The observing night had a few thin cirrus and an average seeing of about 1\farcs5.
The reduction package {\sc specred} of {\sc iraf}\footnote{{\sc iraf} is distributed by the National Optical Astronomy Observatory (NOAO), which is operated by the Association of Universities for Research in Astronomy (AURA) under cooperative agreement with the National Science Foundation (NSF).}
was used to perform the usual reduction process (bias subtraction, spectra extractions and flat-field correction) as well as both wavelength and flux calibrations.

The integral field data are affected by the differential atmospheric refraction (DAR), 
which produces wavelength-dependent offsets on the plane of the sky along the local parallactic angle, with a magnitude that depends on the air mass of the observation and the local atmospheric conditions.
We correct for this effect by implementing the method outlined by \cite{hohenkerksinclair85}\footnote{Reproduced, with permission, from data supplied by HM Nautical Almanac Office, UKHO \copyright\ Crown Copyright.}
in our {\sc idl} routines, which work in a similar way to those written by P.T.~Wallace for {\sc starlink} and \cite{walshroy90}.
The procedure calculates fractional spaxel shifts for each wavelength with respect to a given reference.
We select \hbeta{} and \halpha{} as reference wavelengths for the blue and red spectral ranges, respectively.
The maximum DAR shifts for our observations were 0\farcs59 in right ascension and 1\farcs15 in declination between [\ion{O}{ii}] \wav{3728} and \hbeta{}.
The emission line maps of \hbeta{} and \halpha{} were also used to spatially align the blue and red ranges.
We found a spaxel offset of 0\farcs36 in right ascension by looking for the maximum correlation between both \hbeta{} and \halpha{} maps.
Taking into account the calculated DAR fractional shifts and the alignment offset, the PMAS data were finally corrected using the {\sc bilinear} function of {\sc idl}, which uses a bilinear interpolation algorithm to compute the DAR-corrected data.
The final result of this process reduces the original FoV to \(14\arcsec \times 14\arcsec\).
The DAR correction was tested by checking the alignment of the proplyd peak positions at different wavelengths (e.g.,
\ion{H}{i} lines), finding a good agreement.

The observed field is presented in Fig.~\ref{f1}.
It is located roughly \(25\arcsec\) to the south east of \thC{} at coordinates: \(\mathrm{RA} = \HMSf{5}{35}{17}{4}\) and \(\mathrm{Dec} = \DMS{-5}{23}{41}\).
The field covers interesting morphological structures, in particular several proplyds: 170-334, 170-337, 171-340, 173-341, and 177-341, of which 177-341 is clearly the most prominent.
This field also includes a faint arc of emission in front of 177-341, in the radial direction to \thC{}.
This arc is a bowshock produced by the interaction of the photoevaporated flow from the proplyd ionization front and the fast wind from the main ionizing star \citep[see][and references therein]{henney00}.
Another interesting feature is the Herbig-Haro object HH~514, which begins as a redshifted microjet arising from the proplyd 170-337.
This microjet has a counterjet in the opposite direction, but it is too faint to be detected in even the \textit{HST} images \citep[see][]{ballyetal00, odellhenney08}.
In Fig.~\ref{f1} we have also included the \halpha{} image obtained with PMAS and rebinned to \(160\times 160\) pixels using the {\sc rebin} function of {\sc idl}, which performs an expansion of the original map through a linear interpolation.
We further applied a Gaussian filter to smooth the original \halpha{} image.
With this method, the spatial distribution of certain features observed with the \textit{HST} in our PMAS maps has been nicely reproduced:  three out of five proplyds (173-341 and 171-340 are too faint) and the faint bowshock in front of 177-341.
In the \halpha{} map a flux enhancement can be also located in the northwest corner of the field, which is probably related to the HH~514 object.
However, part of this information is lost following the DAR correction when the FoV is reduced to \(14\arcsec\times 14\arcsec\).
This rebinned and smoothed \halpha{} map is used to construct the \halpha{} contours presented in all maps throughout this paper.
It should be noted that use of these contours on the sample maps produces a slight spatial shift between the maximum contour value and the maximum intensity on the maps.
The \halpha{} contours are mainly useful in illustrating the localization of the various structures.

\subsection{Extraction of proplyd spectra} \label{extrac}
We have extracted several representative spaxels with the aim to analyse in detail the nebular properties of the proplyds detected in our \halpha{} map.
In Fig.~\ref{f1} the red boxes represent the selected spaxels, which coincide with local maxima of \halpha{} emission corresponding to the proplyds 177-341, 170-337 and 170-334, located at the coordinate positions \coords(5:4, -2:-1)\footnote{Hereinafter, the localization of an individual spaxel or multiple spaxels will be given  as \coords(\alpha_1:\alpha_2, \delta_1:\delta_2) according to the coordinate system used in the maps shown in next sections.}, \coords(-4:-5,5:6) and \coords(-5:-6,2:3), respectively.
\cite{ballyetal98,ballyetal00} obtained a wonderful set of narrowband images at different wavelengths ([\ion{O}{iii}], [\ion{O}{i}], \halpha{}, [\ion{N}{ii}] and [\ion{S}{ii}]) for several proplyds, including 170-337 and 177-341.
These images show that the maximum emission line intensity comes from their bright cusps, while the tails are much fainter.
Given the very small angular sizes of the proplyds, it is clear that the observed spectra of the areas of the field we associate with proplyds are only a composite of both proplyd and background emissions.
The largest proplyd is 177-341, with a diameter of about 0\farcs7, but even this is smaller than our spatial resolution and the average seeing during the observations.
Therefore, it is vital to disentangle the intrinsic emission of these objects from that of the ambient ionized gas and other possible contributions \citep{henneyodell99, vasconcelosetal05, tsamisetal11}.
To do that, we have selected a single spaxel as representative for each proplyd, ensuring that it contains the entire or most of the emission from their cusps.
In principle, the main source of contamination is the emission from the background of the nebula, which has to be properly subtracted.
However, it is difficult to define an accurate nebular background since the nebula shows significant brightness variations at small spatial scales.
Furthermore, the opacity due to the dust inside the proplyds is another unknown factor that may introduce systematic errors in the subtraction process of the nebular emission (see \S\ref{dust}).
This issue is extensively discussed in \cite{henneyodell99}, 
who found an effective extinction of \(A_V \simeq 0.1\)--\(0.2\) in the proplyd 177-341, 
which implies effective extinction coefficients for the proplyd, \chb\proplyd, of about 0.04--0.08 dex for their observing conditions and extracted aperture.

In order to put limits on the possible effects of proplyd dust on our extracted spectra, we consider two extreme possibilities.
First, the case where the proplyd lies in front of all the nebular emission and is totally opaque, \(\chb\proplyd \gg 1\), so that the observed (background plus proplyd) emission directly corresponds with the proplyd emission itself.  
Second, the case where the proplyd is totally transparent, \(\chb\proplyd =0\), or lies behind the nebular emission, so that the proplyd spectrum must be obtained by subtracting the full nebular emission from the observed spectrum.
These two cases are particularly easy to deal with since they do not depend on the properties of the dust inside the proplyds. 
The more complex intermediate case of a partially translucent proplyd is deferred until \S\ref{dust}. 

\begin{figure}
  \centering
  \includegraphics[scale=0.5]{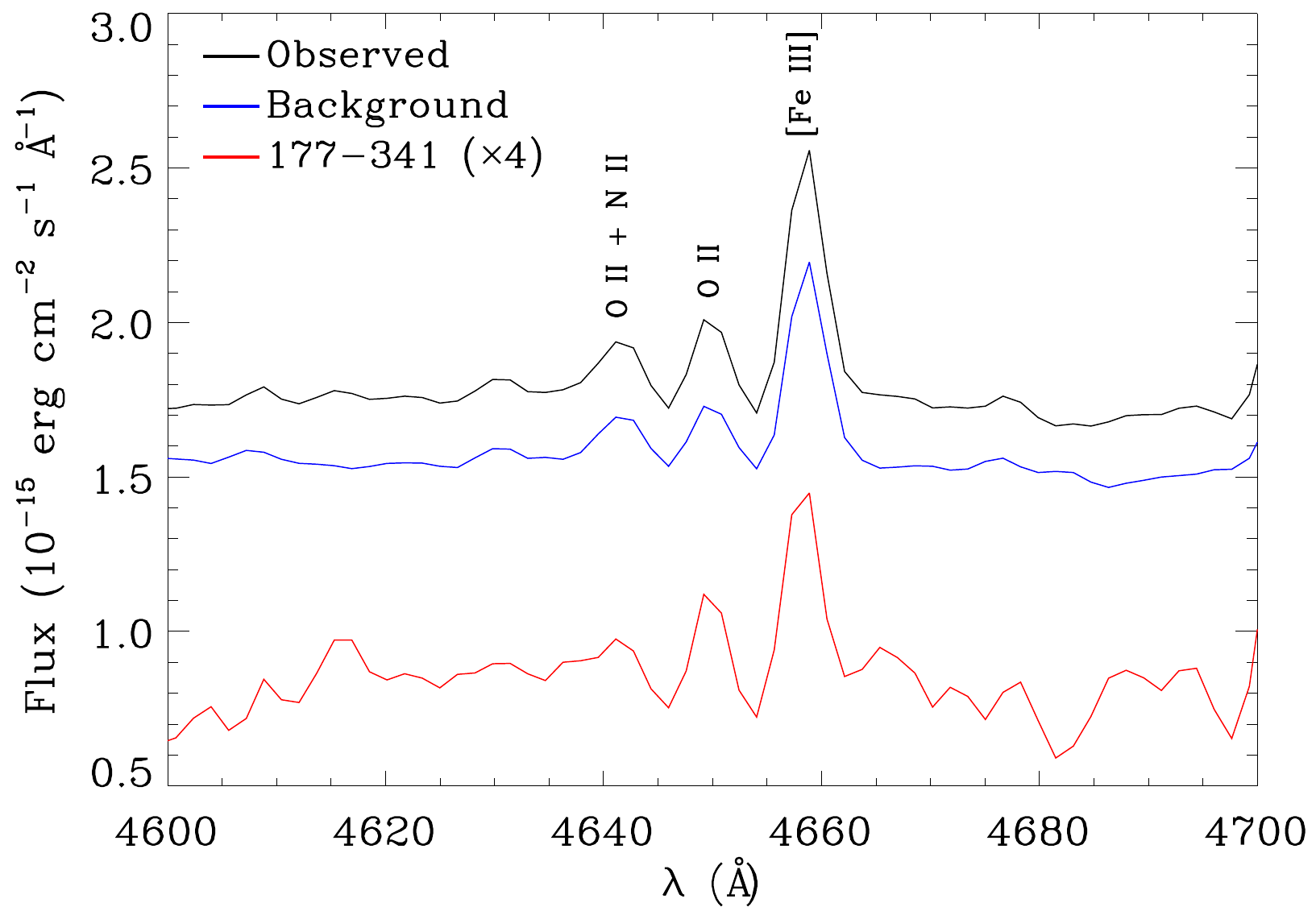} 
  \caption{
    The PMAS spectra of the proplyd 177-341 [spaxel position \coords(5:4,-2:-1)] from 4600 to 4700~\AA, showing the \ion{O}{ii} M1 multiplet lines.
    The black (upper) line shows the observed spectrum (proplyd plus background emission), while the blue (middle) line is the average spectrum of the background emission (see Fig.~\ref{f1}).
    The red (lower) line is obtained as a direct subtraction of the average background emission from the observed one (see \S\ref{extrac}) representing the spectrum of the proplyd 177-341 itself.
    The flux of the proplyd has been multiplied by 4 to facilitate the comparison with the other spectra.}
  \label{f2}
\end{figure}

Taking into account the above considerations, it seems reasonable to obtain the intrinsic proplyd spectra by assuming the transparent case, that is a direct subtraction of the nebular background from the observed spectra (proplyd plus background). 
This is the same approach that was applied in integral field studies of the LV2 proplyd \citep{vasconcelosetal05, tsamisetal11}.
As previously mentioned, the main difficulty with this method lies in defining the nebular background.
As a result, only rough estimations can be obtained from observations assuming an homogeneous background around the proplyds.
In this case, the background emission was obtained as an average spectra from several spaxels around each proplyd.
These spaxels are represented as yellow boxes in Fig.~\ref{f1}.
The proplyds 170-334 and 170-337 shared the three spaxels to the east of 170-337 to construct their associated background spectra.
In order to check the consistency of our results, additional background spectra were obtained from other parts of the field to compare them with the backgrounds assigned to the proplyds.
We estimate that differences among the different backgrounds are lower than 10\%. 
This variation in the background intensity contributes significant uncertainty to the proplyd flux measurements of those lines that are either (1) relatively weak in the proplyd compared with the background (e.g., [\ion{S}{ii}] nebular lines; see \S\ref{lmea}), or (2) relatively faint compared with the adjacent continuum emission (e.g, \ion{C}{ii} or \ion{O}{ii} RLs).

Finally, we also have to consider other possible sources of contamination in the spectra of the selected spaxels, such as the presence of high-velocity emission from any jet associated with the proplyds.
Only for 170-337 has a jet has been reported, HH~514 \citep{ballyetal98, ballyetal00, odellhenney08}, but it is rather faint and its contribution to the total emission should not be important.

\section{Line measurements and reddening correction} \label{medpmas}
 \subsection{Line measurements} \label{lmea}
The flux of each emission line for each individual spaxel in the PMAS field was measured by fitting a Gaussian profile plus a linear continuum using the {\sc splot} routine of {\sc iraf}.
Where necessary, such as for blended lines, multiple Gaussian components were fit simultaneously.  
The brighter lines were measured by an automated process using our own {\sc iraf} scripts, while the fainter ones (e.g.,
\ion{C}{ii} \wav{4267} or \ion{O}{ii} \wav{4650}) were measured by hand in all spaxels.
The emission line fluxes of the individual proplyd spectra (background-subtracted and observed) and of their associated backgrounds were also measured by hand using the same methodology.
The errors in the flux measurements were determined according to the criteria defined by \cite{mesadelgadoetal08}.
The total error for each line was estimated as the quadratic sum of the flux measurement error and the flux calibration error.

   \begin{figure*}
   \centering
   \includegraphics[scale=0.95]{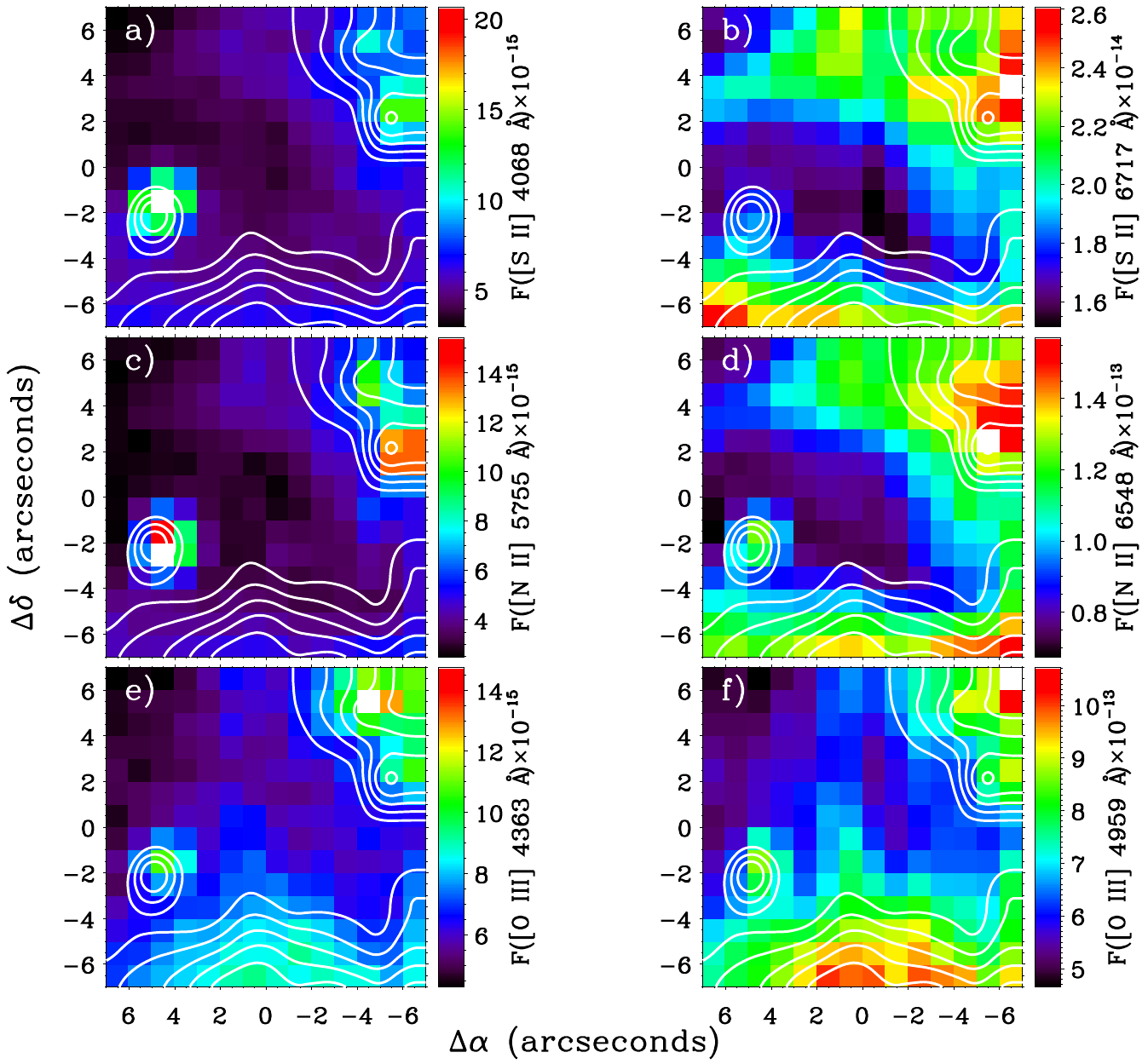} 
   \caption{Spatial distribution maps of auroral (left-hand column) and nebular (right-hand column) emission lines of different ions: a, b) [\ion{S}{ii}] \wav{4068} and \wav{6717}; c, d) [\ion{N}{ii}] \wav{5755} and \wav{6583}; and e, f) [\ion{O}{iii}] \wav{4363} and \wav{4959}. The fluxes are in units of erg~cm$^{-2}$~s$^{-1}$ and \halpha{} contours are overplotted in all maps.}
   \label{f3}
  \end{figure*}
    
In the red spectral range
we found a disagreement between the emission fluxes measured in the short and long exposures.
It is important to correct for this effect because the brightest lines of that range, 
such as \halpha{},
are saturated in the long exposures and must be are measured in the short exposures.
The disagreement is insignificant (lower than the calibration error) in most spaxels covering the nebular background,
but is larger at the proplyd positions.
The measured short-exposure fluxes are on average 11\%, 8\% and 3\% higher than the long-exposure fluxes for 177-341, 170-337 and 170-334, respectively, 
which leads to an even larger discrepancy in the background-subtracted values of 35\%, 29\% and 18\%, respectively.
Since this only affects the point sources within the FoV, it is likely due to seeing variations during the observing night.
Indeed, the effect largely disappears (becoming smaller than the calibration error) if a larger sample region is used to obtain the intrinsic proplyd emission of 177-341. 
We therefore correct for the effect by using the above factors, 
which are rather constant throughout the red spectral range, 
to rescale the short-exposure fluxes. 
For the blue spectral range the differences between the short and long exposures were always lower than the calibration error, and so no correction was necessary.

The emission lines used in our analysis were: \ion{H}{i} Balmer lines, from \halpha{} to H12, 
to compute the reddening correction and test the DAR correction; 
CELs of different ionic species to derive physical conditions and chemical abundances; 
and the faint \ion{C}{ii} and \ion{O}{ii} lines to determine the ionic abundances from RLs.
These lines were easily detected in the whole field, 
excluding the extracted spaxels of the proplyds 170-337 and 170-334, and a few spaxels around them.
The spectrum of the proplyd 177-341 was the only one with high enough signal-to-noise ratio to detect these faint lines in both observed and background-subtracted spectra.
A section of the extracted spectra of this proplyd is presented in Fig.~\ref{f2}, where we can see the blend of \ion{O}{ii} \wav{4649}, \wav{4651} lines before and after the background subtraction. 
 
In Fig.~\ref{f3} we present several representative line flux maps.
The different rows show the spatial distributions of auroral (left-hand column) and nebular (right-hand column) emission lines for the same ion: \ioni{S}{+} in Figs~\ref{f3}(a)--\ref{f3}(b), \ioni{N}{+} in Figs~\ref{f3}(c)--\ref{f3}(d) and \ioni{O}{2+} in Figs~\ref{f3}(e)--\ref{f3}(f).
For all three proplyds, there are significant differences between the spatial distributions of the auroral and nebular lines.
The [\ion{S}{ii}] flux distribution shows the most extreme behaviour, 
whereby the proplyds are barely detectable in the [\ion{S}{ii}] \wav{6717} nebular line map, 
except perhaps for 170-337 (Fig.~\ref{f3}b).
In contrast, they show clear local maxima in the distribution of [\ion{S}{ii}] \wav{4068} auroral line (Fig.~\ref{f3}a).
In fact, for 177-341, we do not detect any trace of [\ion{S}{ii}] \wav{6717} emission in the background-subtracted spectrum.
On the other hand, the spatial distributions of [\ion{N}{ii}] and [\ion{O}{iii}] lines show a somewhat different behaviour for the different proplyds.
The [\ion{N}{ii}] \wav{5755} map has prominent peaks at the positions of the three proplyds (Fig.~\ref{f3}c),
whereas the [\ion{N}{ii}] nebular emission (Fig.~\ref{f3}d) is more extended at the positions of the proplyds 170-334 and 170-337, and has a secondary maximum in 177-341.
For [\ion{O}{iii}], the auroral line map (Fig.~\ref{f3}e) has its maximum at the position of 170-334, a secondary peak at 177-341 and an extended distribution at 170-337.
The spatial distribution of the [\ion{O}{iii}] nebular lines (Fig.~\ref{f3}d) seems to be somewhat more extended at the positions of the three proplyds.
We have found similar behaviour for other emission lines.
For instance, [\ion{O}{i}] \wav{6300} has a spatial distribution very similar to [\ion{N}{ii}] \wav{5755} and [\ion{S}{ii}] \wav{4068} with prominent peaks at the positions of the proplyds.
Another interesting case is the [\ion{O}{ii}] \wav{3728} emission line, 
whose spatial distribution is very similar to that of the [\ion{S}{ii}] nebular line, 
showing the same absence of peaks at the proplyd locations.
Indeed, we did not detect any emission of [\ion{O}{ii}] \wav{3728} line at 177-341 after the background subtraction.
Other emission lines such as [\ion{Ne}{iii}], [\ion{S}{iii}] and [\ion{Ar}{iii}] present spatial distributions similar to those of the [\ion{O}{iii}] and [\ion{N}{ii}] nebular lines.

  \begin{table*}
   \centering
   \begin{minipage}{95mm}
     \caption{
       Critical densities\tfnm{a}, \(n\elec\crit\), of the listed emission lines 
       and the fraction (in \%) of the line fluxes that come from the proplyd itself (background-subtracted spectrum) 
       with respect to the observed spectrum (proplyd plus background emission).}
     \label{necrit}
    \begin{tabular}{lccccc}
     \hline         
     	  & & & \multicolumn{3}{c}{F(proplyd)/F(observed) (\%) }\\   
         Ion & $\lambda$ (\AA) & \(n\elec\crit\) (\cmc) & 177-341 & 170-337 & 170-334\\     
     \hline
       \input{tabla_necrit}

     \hline
    \end{tabular}
    \begin{description}
      \item[\tfnm{a}]For the atomic data of \cite{garciarojasetal09}.
    \end{description}
   \end{minipage}
  \end{table*}
 
In Table~\ref{necrit}, for each of the main CELs we show the ratio of the background-subtracted proplyd flux to the total observed flux in the extracted spaxels (see \S\ref{extrac}),
with the aim of identifying from which component nebular and auroral lines are emitted.
Regarding the emission lines presented in Fig.~\ref{f3}, we see that the [\ion{S}{ii}] auroral lines in the background-subtracted spectra of the proplyd 177-341 represent about 75\% of the observed flux, 
as opposed to only 5\% for the nebular lines.
For [\ion{N}{ii}], the nebular lines are dominated by the background emission (76--88\% of observed flux), 
whereas the proplyd emission dominates (56--77\%) in the auroral line.
Combining the information of Fig.~\ref{f3} and Table~\ref{necrit}, 
it is found that, compared to the Orion background, 177-341 has a much higher surface brightness in the [\ion{S}{ii}] \wav{4068} auroral line than in the [\ion{S}{ii}] \wav{6717} nebular line (factor of \(15\)), 
followed by 170-334 (factor of \(9.8\)) and 170-337 (factor of \(3.2\)).
A similar pattern is seen for both [\ion{N}{ii}] auroral and nebular lines (with more similar factors of \(3.2\), \(2.5\) and \(4.7\), respectively, for the three proplyds).
In contrast to the [\ion{S}{ii}] and [\ion{N}{ii}] lines, where the emission of nebular (auroral) lines from the proplyds is a small (large) fraction of the total observed (proplyd plus background) emission along the line of sight, the [\ion{O}{iii}] auroral and nebular line fraction about 40--50\% and 25\% respectively.
As we will discuss in \S\ref{density}, these patterns can be attributed to collisional de-excitation at the high electron densities expected in the proplyds \citep[up to \(10^6~\cmc\);][]{henneyetal02, tsamisetal11}.

\subsection{Reddening correction}  \label{dchb}
   \begin{figure}
   \centering
   \includegraphics[scale=1.2]{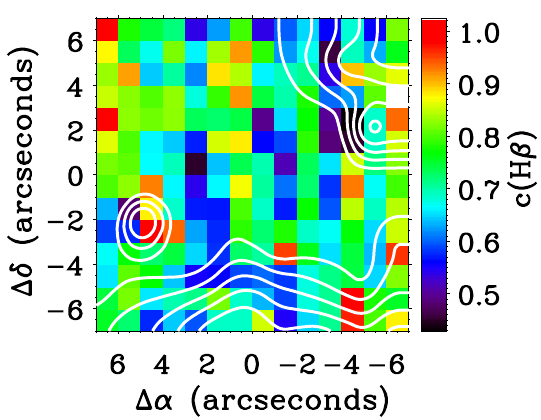} 
   \caption{Spatial distribution map of the extinction coefficient, \chb, along the line of sight with \halpha{} contours overplotted.}
   \label{f4}
  \end{figure}
\begin{table*}\centering 
  \rotatebox{90}{%
    \begin{minipage}{200mm}
      \caption{Detected emission lines used for the analysis of the
        proplyds 177-341, 170-337 and 170-334 for the
        background-subtracted proplyd spectra (\(\chb\proplyd =0\)), the
        observed spectra (\(\chb\proplyd \gg 1\)) and adjacent background
        spectra. All fluxes are dereddened and normalized such that
        \(I(\hbeta) = 100\).}
      \label{flujos}
    \end{minipage}
  }~\rotatebox{90}{%
    \begin{tabular}{cllccccccccc}
      \hline 
      & & & \multicolumn{3}{c}{177-341} & \multicolumn{3}{c}{170-337} &  \multicolumn{3}{c}{170-334} \\  
      $\lambda_0$& Ion & Mult.& \(\chb\proplyd =0\)& \(\chb\proplyd \gg 1\)& Background & \(\chb\proplyd =0\)& \(\chb\proplyd \gg 1\)& 
      Background & \(\chb\proplyd = 0\)& \(\chb\proplyd \gg 1 \)& Background \\
      \hline
      \input{tablalines}
      \hline
    \end{tabular}
  }~\rotatebox{90}{%
    \tfnm{a}Telluric emission is not subtracted from either observed or background spectra.  
  }
\end{table*}

The extinction coefficient, \chb{}, along the line of sight was derived for each spectrum by comparing the observed H$\gamma$/\hbeta{} and H$\delta$/\hbeta{} ratios with the theoretical values predicted by \cite{storeyhummer95} for \(\nel = 1000~\cmc\) and \(\te = \num{10000}~\K\).
Additional \ion{H}{i} line ratios were also measured but they were discarded in deriving the final \chb\ value because they were rather faint (H11 and H12) or were blended with other brighter lines (H7, H8 and H9).
For the background-subtracted proplyd spectra we used the same ratios despite the very high densities expected for the proplyds.
We do not expect significant changes in the results because of the weak density-dependence of the Balmer line ratios.
In fact, the selected ratios show a variation less than 2\%  when electron density and temperature change from \(1000\) to \(10^6~\cmc\) and from \(7500\) to \(\num{12500}~\K\) \citep{storeyhummer95}.
The extinction law normalized to \hbeta{}, $f(\lambda)$, determined by \cite{blagraveetal07} for the Orion Nebula ($R_V=5.5$) was adopted.
This extinction function produces \chb\ coefficients about 0.1~dex higher than the classical one derived by \citep{costeropeimbert70}.
For each spectrum the final \chb\ value was derived as an average of the individual \chb\ determinations from each Balmer line ratio, weighted by their uncertainties, 
giving results with a typical uncertainty in \chb\ of about 0.20~dex.
All flux measurements were referenced to \hbeta{} and \halpha{} for the blue and red range, respectively.
After the reddening correction, we produce a final homogeneous set of flux ratios by re-scaling the flux ratios of the red range to \hbeta{} using the theoretical \halpha{}/\hbeta{} ratio for \(\te = \num{10000}~\K\) and \(\nel = 1000~\cmc\).
In Table~\ref{flujos} we present the dereddened flux ratios of the main emission lines used in our analysis for the extracted proplyd spectra together with their associated backgrounds.

The spatial distribution map of \chb\ along the line of sight is shown in Fig.~\ref{f4}.
The \chb\ coefficient ranges from 0.5 to 1.0 dex, with an average value of 0.74 dex and a standard deviation of 0.14 dex.
This mean value is consistent with the measurements of \cite{odellyusefzadeh00}, 
who derived the \chb\ map from the \halpha{}/\hbeta{} line ratio, finding values of 0.6 to 0.8 at the position of our PMAS field, 
even though a different extinction law was used by these authors.
In Tables~\ref{pro} and \ref{pro2} we present the extinction coefficients along the line of sight, labeled as \(\chb\los\), obtained for the background-subtracted spectra (\(\chb\proplyd =0 \)), the observed spectra (\(\chb\proplyd \gg1 \)) and the adjacent background for each extracted proplyd.
From the extinction map we have estimated mean \chb\ coefficients at the proplyd positions, which amount to 0.80 dex for 177-341, 0.83 dex for 170-337 and 0.80 dex for 170-334.
We find a good agreement for each proplyd between these values and those of Tables~\ref{pro} and \ref{pro2} considering the uncertainties.
The background values also agree with the extinction map of Fig.~\ref{f4}.
Furthermore, our extinction values along the line of sight of the proplyds 177-341 and 170-337 are also consistent with those presented by \cite{odell98}, 0.81 dex and 0.84 dex, respectively.
Finally, we note that the reddening determinations for the background-subtracted and observed spectra of each proplyd are in agreement considering the errors.
This is expected if both emissions arise from the same spatial positions along the line of sight.
However, the small differences between those pairs of values could be the result of internal extinction produced by the presence of dust inside the nebula or the contribution of scattered light.

   \begin{figure*}
   \centering
   \includegraphics[scale=0.95]{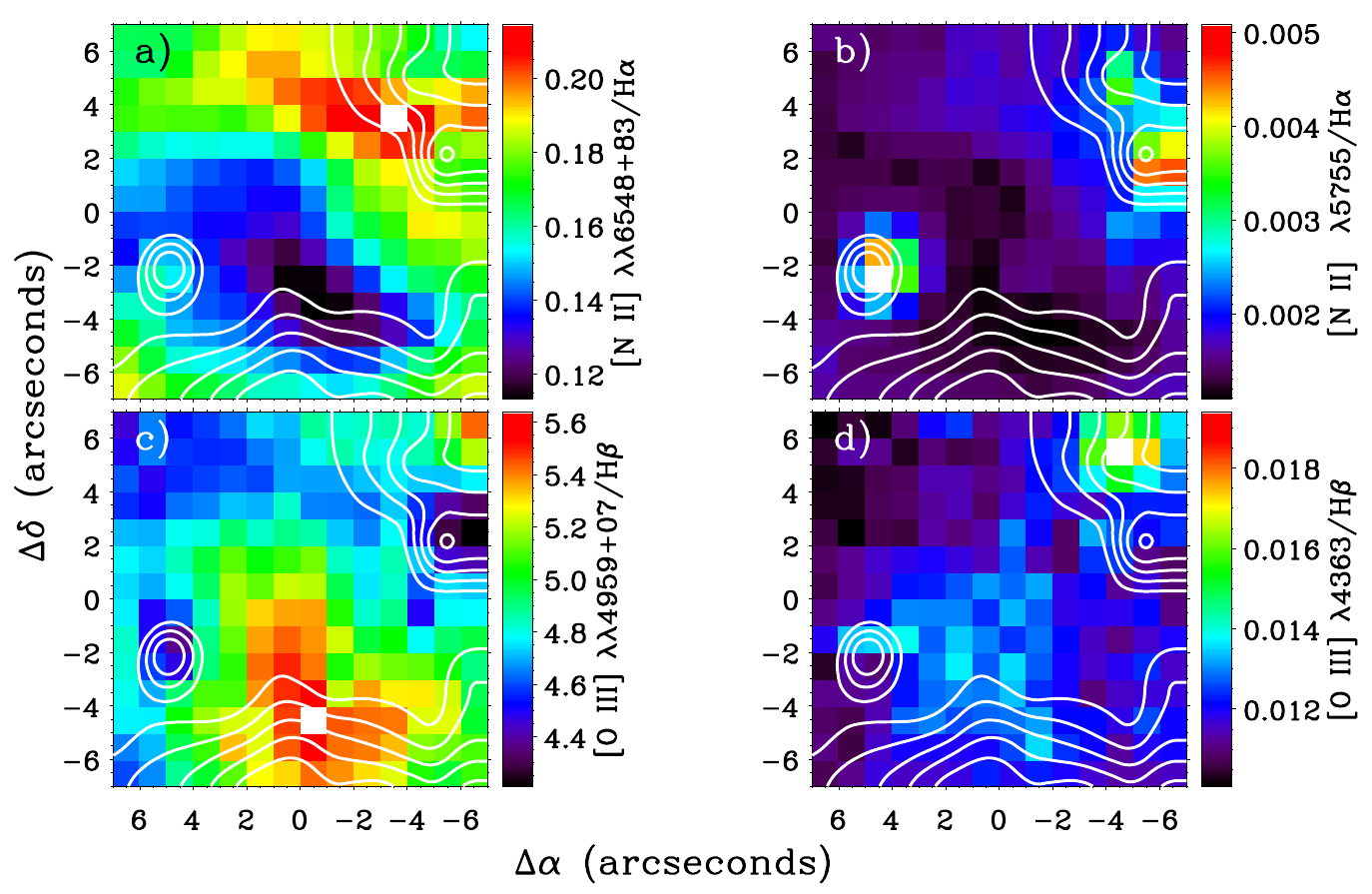} 
   \caption{Spatial distribution maps of the relative line fluxes of nebular (a and c) and auroral (b and d) lines of [\ion{N}{ii}] and [\ion{O}{iii}]. \halpha{} contours are overplotted in all maps.}
   \label{f5}
  \end{figure*}
      
\section{Results and discussion} \label{results}
\subsection{Spatial distributions of line flux ratios} \label{extcoc} 
In this section we explore the behaviour of the [\ion{N}{ii}]/\halpha{} and [\ion{O}{iii}]/\hbeta{} line ratios to study the spatial distribution of the gas excitation.
Assuming that the pattern observed in the spatial distributions of Fig.~\ref{f3} is produced by the effect of collisional de-excitations (see \S\ref{density}), we have plotted in Fig.~\ref{f5} the excitation ratios obtained from nebular and auroral lines.
On the one hand, we show the typical ratios using the [\ion{N}{ii}] \wav{6548}, \wav{6583} and [\ion{O}{iii}] \wav{4959}, \wav{5007} nebular lines (Figs~\ref{f5}a and \ref{f5}c), which allow us to analyse the excitation of the background emission.
On the other hand, we have also used the [\ion{N}{ii}] \wav{5755} and [\ion{O}{iii}] \wav{4363} auroral lines (Figs~\ref{f5}b and \ref{f5}d) to explore the excitation ratios at the proplyd positions, where the nebular lines would be affected by collisional de-excitation effects.
The ionization structure of the background emission seems to be rather similar using both nebular and auroral line ratios.
The low excitation gas is basically surrounding the high excitation areas as we see in Figs~\ref{f5}(a) and \ref{f5}(c).
The maximum emission of the low ionization gas comes from the northwest corner, while the high ionization gas is located at the south side of the field.
This stratification seems to be related to the edge of the Red Bay and some of the Dark Lanes, which are kinematic features identified by \cite{garciadiazhenney07} in this area of the Orion Nebula.
In the case of high ionization gas, some of the higher values seem to be located at the position of the bowshock, spaxels \coords(2:0,-3:0).
At the proplyd positions, the excitation ratios using the nebular lines reveal that the emission from the proplyds 177-341 and 170-337 comes from low ionization gas.
This is clearly confirmed when we use the excitation ratios based on auroral lines, Figs~\ref{f5}(b) and \ref{f5}(d), where we find prominent spikes at these positions.
It is an expected behaviour because we are observing the radiation emitted from the low ionization gas in the ionization front of the cocoon structure assumed for the proplyds \citep[see figure 5 of][]{henneyodell99}.
The proplyd 170-334 also has low ionization emission, but the emission of auroral lines of high ionization gas is dominant for this proplyd at spaxels \coords(-4:-3,5:6).
This behaviour may imply a extremely high density for the proplyd 170-334, which is quenching the emission of the [\ion{O}{iii}] \wav{4959}, \wav{5007} nebular lines (see \S\ref{density}).

\subsection{Physical conditions} \label{phycon}
\begin{figure*}
   \centering
   \includegraphics[scale=0.95]{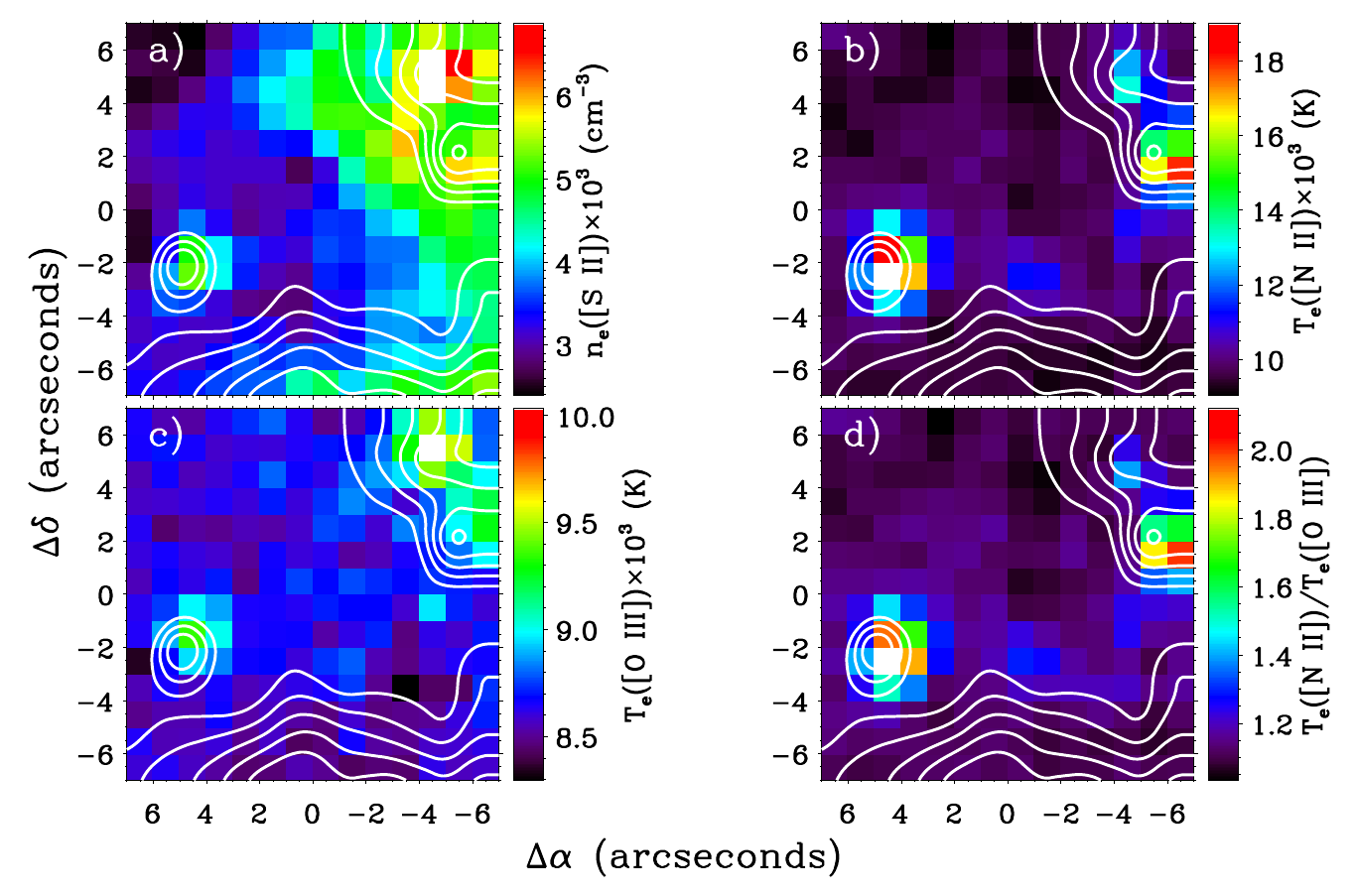} 
   \caption{Spatial distribution maps of physical conditions with \halpha{} contours overplotted: a) \nel([\ion{S}{ii}]), b) \te([\ion{N}{ii}]), c) \te([\ion{O}{iii}]) and d)  \te([\ion{N}{ii}])/\te([\ion{O}{iii}]) ratio.}
   \label{f6}
  \end{figure*}

In this section we present electron densities, \nel, and temperatures, \te, derived from the analysis of each spaxel as well as the extracted spectra of the proplyds.
The spatial distribution maps of \te\ and \nel\ are shown in Fig.~\ref{f6} and the physical conditions derived for the background-subtracted, observed and adjacent background spectra of each proplyd are available in Tables~\ref{pro} and \ref{pro2}.
In Table~\ref{pro} we also present the different results of the proplyd 177-341 by taking into account the effect of opacity in the proplyd on the background subtraction procedure (see \S\ref{dust}). Here the opacity is parametrized by \(\chb\proplyd  \).

The calculations for each spaxel were performed using the {\sc temden} task of the {\sc nebular} package \citep{shawdufour95} of {\sc iraf} with updated atomic data \citep[see table 6 of][]{garciarojasetal09}.
An iterative procedure was used to determine \te\ and \nel{} from the usual CEL diagnostic ratios: [\ion{S}{ii}] \wav{6717}/\wav{6731} for \nel, and [\ion{O}{iii}] \wavs{4959}{5007}/\wav{4363} and [\ion{N}{ii}] \wavs{6548}{6583}/\wav{5755} for \te.
We assume an initial \(\te = \num{10000}~\K\) to derive a first estimate foe \nel([\ion{S}{ii}]), then \te([\ion{O}{iii}]) and \te([\ion{N}{ii}]) are calculated.
The process is repeated until convergence, which is reached after 3 or 5 iterations, taking \te([\ion{N}{ii}]) as representative for \ioni{S}{+} ion assuming a two-zone scheme (see \ref{cheab}).

\begin{figure*}
  \centering

  \includegraphics[scale=0.4]{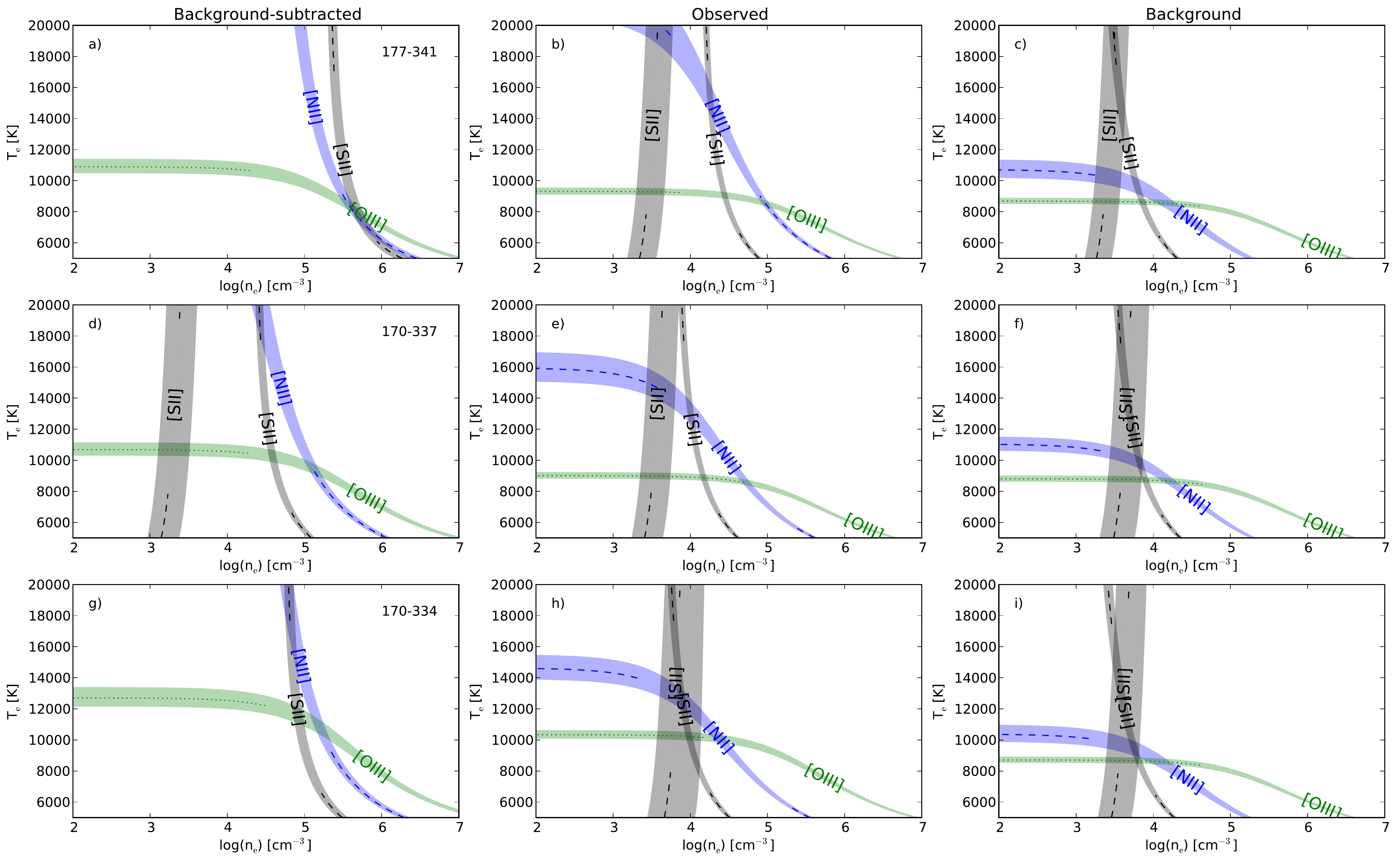} 
  \caption[captionlarge]{\nel--\te\ diagnostic curves for the background-subtracted, observed and adjacent background spectra for the extracted proplyds: 177-341 (first row), 170-337 (second row) and 170-334 (third row). The different curves represent the following diagnostic ratios: [\ion{N}{ii}] \wav{6583}/\wav{5755}, [\ion{O}{iii}] \wav{5007}/\wav{4363}, [\ion{S}{ii}] \wavs{6717}{6731}/\wavs{4068}{4076} (narrow band) and  [\ion{S}{ii}] \wav{6717}/\wav{6731} (wide band). The width of each curve represents the diagnostic uncertainty.}
  \label{f61}
\end{figure*}

The physical conditions given in Tables~\ref{pro} and \ref{pro2} for the extracted spectra of the proplyds (background-subtracted and observed) as well as their adjacent backgrounds were computed following a similar iterative procedure.
Additionally, we use alternative density diagnostics that are more suitable for deriving the expected high electron densities for the proplyds, which can reach values exceeding \(10^6~\cmc\) \citep{henneyetal02, tsamisetal11}.
At those densities, diagnostics based on typical optical line ratios 
(e.g., [\ion{S}{ii}] \wav{6717}/\wav{6731} or [\ion{Cl}{iii}] \wav{5517}/\wav{5537}) 
would not return reliable values due to the low critical densities of the upper levels of the transitions producing those lines.
The first alternative density indicator is based on the analysis of [\ion{Fe}{iii}] emission lines, which are useful as robust density diagnostics with much wider ranges of validity than other line ratios.
As shown in Table~\ref{flujos}, we have detected several [\ion{Fe}{iii}] lines (a total number of 5 lines of the 3F multiplet and one of the 2F multiplet) in the background-subtracted spectra of 177-341, in the observed spectra of 177-341 and 170-337, and in the three adjacent background spectra of each proplyd.
The density adopted from the analysis of [\ion{Fe}{iii}] lines, \nel([\ion{Fe}{iii}]), is the value that provides the minimum dispersion between the observed and theoretical line ratios with respect to [\ion{Fe}{iii}] \wav{4658} for the whole set of lines.
The theoretical [\ion{Fe}{iii}] emissivities were computed by solving a 34-level atomic model using the atomic data of \cite{zhang96} for collisional strengths, and \cite{quinet96} and \cite{johanssonetal00} for transition probabilities.
The second alternative density indicator consists in obtaining diagnostic curves in the plane \nel--\te\ from the line ratios [\ion{N}{ii}] \wav{6583}/\wav{5755}, [\ion{O}{iii}] \wav{5007}/\wav{4363}, [\ion{S}{ii}] \wavs{6717}{6731}/\wavs{4068}{4076} and  [\ion{S}{ii}] \wav{6717}/\wav{6731}.
These diagnostic curves (Fig.~\ref{f61}) were computed making use of {\sc pyneb} \citep{luridianaetal11}, 
which is an updated version of the {\sc nebular} package written in {\sc python} programming language,
and the set of atomic data presented in \cite{garciarojasetal09}.
In general, the intersection of those curves yields a clear solution for \nel--\te, whose values are labeled as diag.
\nel--\te\ in Tables~\ref{pro} and \ref{pro2}.
In particular, we have chosen the values provided by the [\ion{N}{ii}] and [\ion{O}{iii}] curves.
The [\ion{S}{ii}] curves obtained for the background-subtracted and observed spectra of the proplyds require a more detailed discussion because apart from collisional de-excitation of the [\ion{S}{ii}] nebular emission lines, 
uncertainties in the background subtraction could be affecting the results (see next section).
Finally, in the cases where both previous density indicators were available for the background-subtracted and observed proplyd spectra, the adopted values were calculated as weighted averages, otherwise the result from the diagnostic curves was adopted.

For the adjacent background spectra it was not necessary to use the diagnostic diagrams and the densities derived from the [\ion{Fe}{iii}] and [\ion{S}{ii}] lines were weighted to obtain the average final values. 
Then, \te([\ion{O}{iii}]) and \te([\ion{N}{ii}]) were calculated using these adopted densities.
We did not correct for the recombination contribution to the flux of the [\ion{N}{ii}] \wav{5755} auroral line in any of the above calculations.
For the background emission and the observed spectra this is expected to be rather low because of the ionization degree of the Orion Nebula \citep[e.g.][]{estebanetal04} and even less important in the case of the proplyds.

\begin{table*}\centering
  \rotatebox{90}{%
    \begin{minipage}{220mm}
      \caption{Physical conditions and chemical abundances\tfnm{a} for the background-subtracted spectra (\(\chb\proplyd =0 \)), the observed spectra (\(\chb\proplyd \gg1 \)) and the adjacent background of the proplyd 177-341. The same quantities are presented considering the possible opacity of this proplyd based on the toy-model of \S\ref{dust}.}
      \label{pro}
    \end{minipage}
  }~\rotatebox{90}{%
    \begin{tabular}{clccccccccc}
      \hline                                              
      & & \multicolumn{8}{c}{Proplyd Emission} & \multirow{2}{*}{Background}\\  
      \cline{3-10}
      Proplyd dust\multirow{2}{*}{ \rule{0.2pt}{0.67cm}}& $R_V\proplyd$ &  & \multicolumn{2}{c}{$3.1$} & %
      \multicolumn{2}{c}{$5.5$} & \multicolumn{2}{c}{$7$} && \multirow{2}{*}{Emission}\\  
      properties &\(\chb\proplyd  \) & 0 & 0.05 & 0.1 & 0.05 & 0.1 & 0.05 & 0.1 & $\gg$1 &\\
      \multicolumn{2}{c}{\chb\los} & 1.05$\pm$0.30 & 0.82$\pm$0.27 & 0.82$\pm$0.26 & %
      0.76$\pm$0.27 & 0.73$\pm$0.27 & 0.74$\pm$0.27 & 0.71$\pm$0.26 & 0.87$\pm$0.15 & %
      0.82$\pm$0.15\\   
      \multicolumn{11}{c}{}\\                                                                                                          
      \multicolumn{11}{c}{Physical conditions}\\   
      \input{tablaproplyd}

      \hline
    \end{tabular}
    }~\rotatebox{90}{%
      \begin{minipage}{1.0\linewidth}
        \begin{description}
        \item[$^a$] In units of \(12+\log(\ioni{X}{+i}/\ioni{H}{+}\)).
        \item[$^b$] Determinations based on diagnostic curves
          (Fig.~\ref{f61}).
        \item[$^c$] Proplyd emission: weighted average of
          [\ion{Fe}{iii}] and diag. \nel--\te\ densities.  Background
          emission: weighted average of [\ion{Fe}{iii}] and
          [\ion{S}{ii}] densities.
        \item[$^d$] Abundances determined from RLs.
        \end{description}
      \end{minipage}
    } 
   \end{table*}

  \subsubsection{Electron densities} \label{density}
In the spatial distribution map of \nel([\ion{S}{ii}]) shown in Fig.~\ref{f6}(a), we do not find sharp density contrast between the proplyds and background.
The typical values range from \(3000\) to \(8000~\cmc\) with uncertainties between \(1000\) and \(2000~\cmc\).
On average the nebular background has a density of about \(3700~\cmc\) ranging from \(2800\) to \(5000~\cmc\).
The maximum value of \nel([\ion{S}{ii}]) (\(\simeq 8000~\cmc\)) is reached at the position of the proplyd 170-334 and secondary peaks of about \(5800\) and \(5200~\cmc\) are found at the proplyds 170-337 and 177-341, respectively.
As is expected, these values are in agreement within the uncertainties with \nel([\ion{S}{ii}]) values shown in Tables~\ref{pro} and \ref{pro2} for the observed extracted spectra of the proplyds: \(3700~\cmc\) for 177-341, \(4240~\cmc\) for 170-337 and \(8540~\cmc\) for 170-334.
However, the density determinations obtained from [\ion{S}{ii}] line ratios clearly disagree with those derived from the [\ion{Fe}{iii}] lines and the diagnostic curves also included in Tables~\ref{pro} and \ref{pro2}.
The two last indicators point out the existence of high-density ionized gas at the proplyd positions, where the values derived for the observed spectra (proplyd plus background) now amount to \(\num{93100}~\cmc\) for 177-341, \(\num{57900}~\cmc\) for 170-337 and \(\num{24000}~\cmc\) for 170-334.

We interpret this disagreement as due to the fact that the [\ion{S}{ii}] \wav{6717}, \wav{6731} emission detected in the observed spectra of the proplyds comes mainly from the nebular background.
We saw in Table~\ref{necrit} that the proplyd emission itself only represents 4--17\% of the total observed emission of the [\ion{S}{ii}] nebular lines.
If the high densities derived from the [\ion{Fe}{iii}] lines and the diagnostic curves correctly reflect the proplyd physical conditions, then the [\ion{S}{ii}] line ratio will not provide reliable densities because collisional de-excitation will be severely suppressing the emission of [\ion{S}{ii}] nebular lines emitted from these objects due to the low critical densities of the upper levels (see Table~\ref{necrit}).
Therefore, the \nel([\ion{S}{ii}]) values determined for the observed spectra are more similar to the background ones and this can be seen in Tables~\ref{pro} and \ref{pro2}.
This also explains the behaviour of the [\ion{S}{ii}] \wav{6717}/\wav{6731} curves of the observed spectra of the proplyds (Figs~\ref{f61}b, \ref{f61}e and \ref{f61}h), which are similar to those of the background emission (Figs~\ref{f61}c, \ref{f61}f and \ref{f61}i).
Another explanation for this would be that the [\ion{S}{ii}] nebular lines could partly originate from a lower density envelope surrounding the proplyds, but our observations do not have enough spatial resolution to confirm this; however, the generic Orion emission is most probably the dominant contributor to these lines (see Table~\ref{necrit}).
Then, the discussion above clearly indicates that the [\ion{S}{ii}] \wavs{6717}{6731}/\wavs{4068}{4076} diagnostic curves are unsuitable for deriving densities from the observed spectra of the proplyds because the constituent lines of the aforementioned [\ion{S}{ii}] nebular to auroral diagnostic ratio arise in two very different gas volumes (generic nebula vs.
proplyd).
Thus in this case the ratio has a very limited diagnostic value.
In the same way as for the [\ion{S}{ii}] nebular lines, collisional de-excitation will affect very strongly the flux of the [\ion{O}{ii}] \wav{3728} doublet due to its similarly low critical densities (see Table~\ref{necrit}).

The situation discussed in the previous paragraph hampers the correct interpretation of the background-subtracted spectra.
The [\ion{S}{ii}] \wav{6717}, \wav{6731} and [\ion{O}{ii}] \wav{3728} emission should partially or totally disappear after subtracting the background emission from the observed spectra.
However, these lines do not completely disappear in the background-subtracted spectra of 170-337 as we see in Table~\ref{flujos}.
In fact, the relative contribution of the proplyd to the [\ion{S}{ii}] and [\ion{O}{ii}] emission is larger for this object than for the others (see Table~\ref{necrit}).
Furthermore, we derive a density of about \(2300~\cmc\) for 170-337 using the [\ion{S}{ii}] line ratio, which is actually more consistent with the background values and very different from the high density determined from the diagnostic curves, \(1.29 \times 10^5~\cmc\).
This is also the behaviour that we observe in the diagnostic diagram obtained for the background-subtracted spectra of this proplyd (Fig.~\ref{f61}d), where the [\ion{S}{ii}] curves provide lower densities than the intersecion of [\ion{N}{ii}] and [\ion{O}{iii}].
These results indicate that there is residual background contamination in the background-subtracted spectra of this proplyd, probably due to its small size.
We expect a similar situation for the proplyd 170-334, which is even smaller than 170-337.
We do not detect [\ion{S}{ii}] \wav{6717} in the background-subtracted spectrum of this object, but there is some emission from [\ion{O}{ii}] \wav{3728} (see Table~\ref{flujos}) and the curves of [\ion{S}{ii}], [\ion{N}{ii}] and [\ion{O}{iii}] do not have a common intersection (Fig.~\ref{f61}g).
Hence, the physical conditions of these proplyds should be considered with some caution because they could be contaminated by background emission.
On the other hand, background contamination seems to not be substantial in the background-subtracted spectra of 177-341 due to its larger size, the optimal signal-to-noise relation of its spectra, and the appropriate isolation of the selected spaxel.
The plasma diagnostic of this proplyd (Fig.~\ref{f61}a) produces consistent physical conditions for the different curves.
All these considerations are relevant for the analysis of the proplyd properties and, in particular, they have to be considered in the calculations of the chemical abundances from CELs, as we will see in \S\ref{cheab}.

It is also interesting to compare the densities of the background and proplyds derived from the [\ion{Fe}{iii}] analysis and the diagnostic curves (Tables \ref{pro} and \ref{pro2}).
We see that proplyd densities are one to two orders of magnitude higher than the background ones, a similar result to that found for the LV2 proplyd \citep{tsamisetal11}.
For instance, the proplyd 177-341 has densities of about \(3.8 \times 10^5~\cmc\) and \(9.3 \times 10^4~\cmc\) determined from the background-subtracted and observed spectra, respectively, whilst the background density amounts to \(3800~\cmc\).
For the other two objects, 170-337 and 170-334, we find a similar contrast, although the proplyd densities are lower than those of 177-341, perhaps because they are  more affected by background contamination as we mentioned in the previous paragraph. 

In their study of the LV2 proplyd, \cite{tsamiswalsh11} found a clear correlation between the critical density and the relative contribution of the proplyd emission to the total observed one in the area occupied by the object.
In our case, such correlation can be seen in Table~\ref{necrit}, although it is not so clear as in LV2.
However, the densities of our proplyds are different and the background contamination can also affect the smaller proplyds.
The general behaviour of the proplyds observed in our PMAS field shows that emission lines emitted by ions with high ionization potentials (\ioni{O}{2+}, \ioni{Ne}{2+}, \ioni{S}{2+} and \ioni{Ar}{2+}) are dominated by the background emission,
whereas the proplyd contribution to emission lines arising from ions with low ionization potentials (\ioni{N}{+}, \ioni{S}{+} and \ioni{O}{+}) varies depending on the transition type: background emission dominates the nebular lines and proplyd emission dominates the auroral lines.

Finally, inspection of Table~\ref{necrit} also gives further insights into the effect of collisional de-excitation due to high densities in the different spaxels containing proplyd emission.
In all cases, we see a very strong or even total suppression of the [\ion{S}{ii}], [\ion{O}{ii}] and  [\ion{N}{ii}] nebular lines.
This indicates that the true electron densities must be above the critical density of the upper levels that produce those lines.
However, 170-334 presents a somewhat different collisional de-excitation pattern because the contribution of the proplyd to the flux of [\ion{O}{iii}] auroral line is higher (see Table~\ref{necrit} and Fig.~\ref{f3}e).
In addition, the spatial distribution maps of [\ion{O}{iii}]/\hbeta{} ratio (Fig.~\ref{f5}d) and \te([\ion{O}{iii}]) (Fig.~\ref{f6}c) show a clear increase with respect to the surroundings at the position of this proplyd.
These two facts indicate that the true electron density in this proplyd should be even higher than the critical density of the nebular [\ion{O}{iii}] lines, i.e., higher than \(7\times10^5~\cmc\), much larger than the value quoted for the proplyd in Table~\ref{pro2} of about \(1.6\times10^5~\cmc\).
Indeed, the models of \cite{henneyarthur98} give a peak density for 170-334 (erroneously named 171-334 in their Table~2) of about \(3\times10^6~\cmc\), which is 5 times higher than that predicted for 177-341 (\(6.2\times 10^5~\cmc\)).
This is consistent with the [\ion{O} {iii}] nebular lines being collisionally de-excited in 170-334.
Probably, the aforementioned background contamination is producing the artificial low density value presented in Table~\ref{pro2}.

  \subsubsection{Electron temperatures} \label{temperature}
If we know the true representative \nel\ of the ionized gas in the proplyds, then we can derive the appropriate \te\ because statistical equilibrium equations take into account the collisional de-excitation effects.
However, this is not so simple because the physical conditions in a proplyd could be far from homogeneous and, in particular, the density is expected to show significative structure at spatial scales much smaller than can be resolved with our observations.
Evidence of such effects was provided by  \cite{tsamiswalsh11} who combined high-spectral resolution VLT observations of optical [\ion{O}{iii}] nebular and auroral lines with the \textit{HST} far-UV \ion{C}{iii}] observations by \cite{henneyetal02} to simultaneously determine the electron temperature and density in the \ioni{O}{2+} zone of LV2: the temperature shows a maximum root-mean-square variation of 10\% and the density varies by up to 50\% across the line profile associated with the kinematic core of the proplyd.

Assuming that the adopted densities from [\ion{Fe}{iii}] lines and the diagnostic curves are the correct, we determine \te([\ion{N}{ii}]) and \te([\ion{O}{iii}]) for both background-subtracted and observed spectra of the proplyds (Tables~\ref{pro} and \ref{pro2}) and compare these results with the spatial distribution maps calculated using the \nel([\ion{S}{ii}]) map of Fig.~\ref{f6}(a).
The background shows almost constant temperatures, 9800~K for \te([\ion{N}{ii}]) and 8650~K for \te([\ion{O}{iii}]), with typical errors of about 600 and 200~K, respectively.
In both temperature distribution maps (Figs~\ref{f6}b and \ref{f6}c) the proplyds show very prominent spikes, which are more evident in the \te([\ion{N}{ii}]) distribution in the cases of 177-341 and 170-337 and in \te([\ion{O}{iii}]) for 170-334.
It is clear that these peaks do not correspond to the true temperatures of the proplyds because the density used to derive the temperature maps, \nel([\ion{S}{ii}]), is not reliable for the proplyds as we previously commented (\S\ref{density}).
In Tables~\ref{pro} and \ref{pro2}, where we have used the higher densities given by the other indicators, we can see that the resultant temperatures do not show such high contrast.

For 177-341, \te([\ion{O}{iii}]) and \te([\ion{N}{ii}]) are lower than the background temperatures by only a few hundred~K and about 1000~K, respectively.
Interpretation of the temperatures in the proplyds 170-337 and 170-334 is more difficult because of the background subtraction problems as we previously mentioned (see \S\ref{density}).
High temperatures derived from the background-subtracted spectra of these proplyds are probably due to errors in correcting for collisional de-excitation effects occasioned by the adopted density derived from the [\ion{Fe}{iii}] lines and diagnostic diagrams being lower than the true value.
This effect is clear for the proplyd 170-334, where the temperatures are even higher than those of the nebular background.

It is likely that much of the apparent structure seen in temperature indicator maps, especially \te([\ion{N}{ii}]) (Fig.~\ref{f6}b), is in reality due to high-density ionized gas.
In the spatial distribution of \te([\ion{N}{ii}]) we find two secondary maxima of about \SI{11500}{K} at the spaxel positions \coords(0:-2,-3:-2) and \coords(-4:-5,-2:0).
These positions do not correspond to any noticeable morphological structures in our emission line maps (see Fig.~\ref{f3}),
but they do coincide with the positions of the faint proplyds 173-341 and 171-340 (see the \textit{HST} image of Fig.~\ref{f1}).
Careful examination of the spatial distribution of the [\ion{N}{ii}] \wav{5755} line (Fig.~\ref{f3}c) reveals a slight increase in the line flux at these spaxels of about 13\% and 20\%, respectively, with respect to the adjacent background.
On the other hand, no similar enhancements were detected in the spatial distribution of the [\ion{N}{ii}] \wav{6548} nebular line (Fig.~\ref{f3}d).
Therefore, \te([\ion{N}{ii}]) is again overestimated in our detected proplyds because the effect of collisional de-excitation is not well corrected for due to the adoption of the density derived from the [\ion{S}{ii}] line ratio.

   \begin{table*}
   \centering
   \begin{minipage}{160mm}
     \caption{Physical conditions for the background-subtracted spectra (\(\chb\proplyd =0 \)), the observed spectra (\(\chb\proplyd \gg1 \)) and the adjacent background of the proplyds 170-337 and 170-334.}
     \label{pro2}
    \begin{tabular}{clcccccc}
     \hline 
        & & \multicolumn{3}{c}{ 170-337} & 
        \multicolumn{3}{c}{ 170-334 }\\
        \hline
        & & \multicolumn{2}{c}{Proplyd Emission} & Background &\multicolumn{2}{c}{Proplyd Emission}& 
         Background\\  
         \multicolumn{2}{c}{\chb\proplyd} & 0 & $\gg$1 & Emission &0 & $\gg$1 & Emission\\
       \multicolumn{2}{c}{\chb\los} & 0.76$\pm$0.22 & 0.68$\pm$0.15 & 0.72$\pm$0.15 & %
       0.53$\pm$0.28 & 0.70$\pm$0.15 & 0.66$\pm$0.15\\   
       \multicolumn{8}{c}{}\\                                                                                                          
       \multicolumn{8}{c}{Physical conditions}\\   
        \input{tablaoprop}
     \hline
    \end{tabular}
    \begin{description}
      \item[$^a$] Determinations based on diagnostic curves (Fig.~\ref{f61}).  
      \item[$^b$] Proplyd emission: weighted average of [\ion{Fe}{iii}] and diag. \nel-\te\ densities where possible. Background emission: weighted average of [\ion{Fe}{iii}] and [\ion{S}{ii}] densities where possible. 
    \end{description}
   \end{minipage}
  \end{table*}

 \subsection{Chemical abundances from CELs}  \label{cheab}
Making use of the {\sc ionic} task of the {\sc nebular} package \citep{shawdufour95} and the set of atomic data presented in Table~6 of \cite{garciarojasetal09}, we have derived ionic abundances for several ions from CELs: \ioni{N}{+}, \ioni{O}{+}, \ioni{O}{2+}, \ioni{S}{+}, \ioni{S}{2+}, \ioni{Ne}{2+} and \ioni{Ar}{2+}.
We have assumed a two-zone scheme with the absence of temperature fluctuations, \tf = 0, adopting \te([\ion{N}{ii}]) for ions with low ionization potentials (\ioni{N}{+}, \ioni{O}{+} and \ioni{S}{+}) and \te([\ion{O}{iii}]) for ions with high ionization potentials (\ioni{O}{2+}, \ioni{S}{2+}, \ioni{Ne}{2+} and \ioni{Ar}{2+}).
We have also determined the \ioni{Fe}{2+} abundances for the extracted spaxels using \te([\ion{N}{ii}]) and the atomic model described in \S\ref{phycon}.
The abundance uncertainties are the quadratic sum of independent contributions from \nel, \te, and line flux errors.

\begin{figure*}
   \centering
   \includegraphics[scale=0.95]{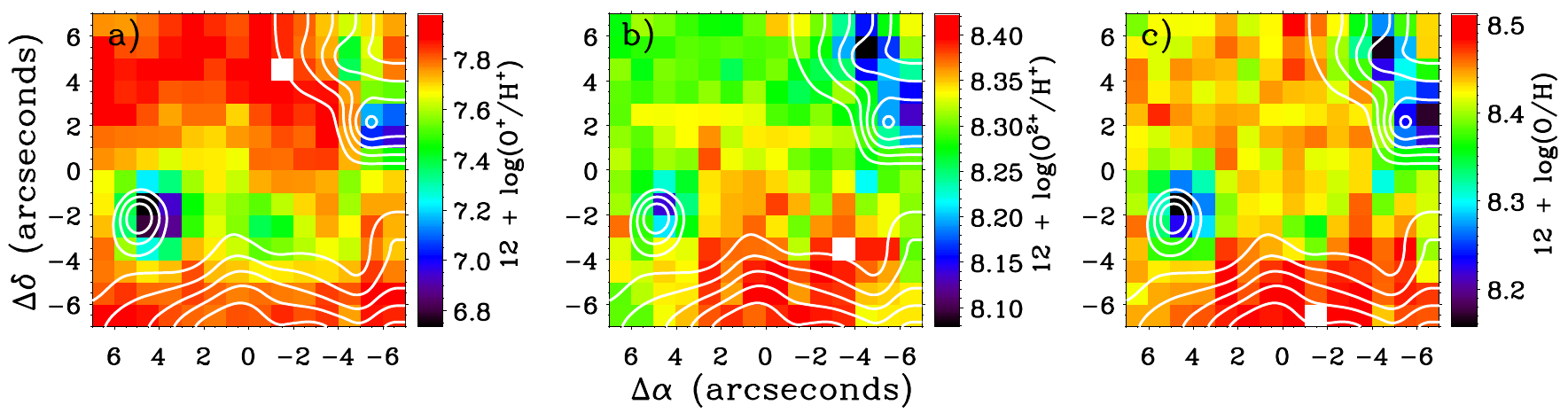} 
   \caption{Spatial distribution maps of (a) \(\ioni{O}{+}/\ioni{H}{+}\), (b) \(\ioni{O}{2+}/\ioni{H}{+}\) and  (c) \(\mathrm{O/H}\) derived from CELs, in units of \(12 + \log(\ioni{X}{+i}/\ioni{H}{+})\). \halpha{} contours are overplotted in all maps.}
   \label{f7}
  \end{figure*}

In Fig.~\ref{f7}, we present the spatial distribution maps of the abundances of two relevant ions, \ioni{O}{+} and \ioni{O}{2+}, as well as of the total O abundance calculated as the sum of \ioni{O}{+} and \ioni{O}{2+} abundances.
The typical uncertainties for each spaxel are about 0.15~dex for \(\ioni{O}{+}/\ioni{H}{+}\), 0.04~dex for \(\ioni{O}{2+}/\ioni{H}{+}\) and 0.10~dex for \(\mathrm{O/H}\).\
In all maps, the minimum values are clearly located at the proplyd positions and their surroundings.
Considering the results discussed in \S\ref{phycon}, this must be related to the overestimation of temperatures due to the use of an unreliable density for the proplyds.
Excluding the proplyd emission areas, the mean value for the total O abundance amounts to \(8.45 \pm 0.05\)~dex, which is in agreement with the previous determinations for the Orion Nebula available in the literature \citep[8.50 dex; e.g.][]{estebanetal04}.
In the spatial distribution of the \ioni{O}{+}/\ioni{H}{+} ratio (Fig.~\ref{f7}a), the values of about 7.6 dex slightly to the south of the center of the field correspond to the areas of higher \te([\ion{N}{ii}]) associated with the proplyd 173-341 at spaxels \coords(0:-2,-3:-2) and the proplyd 171-340 at \coords(-4:-5,-2:0).
In Fig.~\ref{f7}(c), the total O/H ratio also shows secondary minima at those positions where we found the \te([\ion{N}{ii}]) and \te([\ion{O}{iii}]) peaks for the proplyds.
The particular structure of the high-excitation bowshock in front of 177-341, spaxels (2:0,$-$3:0), does not show any noticeable behaviour in the oxygen abundance maps compared with its surroundings.
We measure a total O abundance of about 8.45 dex, similar to the background average.

It is interesting to compare the abundance distribution maps with the values that are derived from the extracted spectra of the proplyds.
For the proplyds 170-337 and 170-334, chemical analysis reveals a complex abundance pattern because of the background contamination and the use of unreliable densities and temperatures.
We have omitted the chemical abundances derived for these proplyds in Table~\ref{pro2} and, hence, focused on the results for 177-341, whose ionic abundances are presented in Table~\ref{pro}.
We also show results considering the effects of opacity in this proplyd (see \S\ref{dust}).
Taking into account the role of the collisional de-excitation mentioned in \S\ref{lmea} and \S\ref{density}, it is clear that the flux of the [\ion{O}{ii}] \wav{3728} doublet in the observed spectra does not provide reliable abundances.
This doublet is not detected in the background-subtracted spectra of 177-341.
Thus, the \ioni{O}{+}/\ioni{H}{+} and O/H ratios have also been omitted in Table~\ref{pro}.
We obtained an \ioni{O}{2+} abundance of about 8.36~dex from CELs for the observed spectra, similar to the adjacent background value, 8.30 dex, and higher than the abundance obtained from the \ioni{O}{2+}/\ioni{H}{+} ratio map (Fig.~\ref{f7}b).
Subtraction of the background emission yields an \ioni{O}{2+}/\ioni{H}{+} ratio of 8.42~dex for 177-341.
The other ions show larger differences in abundance when comparing the observed, background and background-subtracted spectra of the proplyd.
The largest differences with respect to the adjacent background are found for the \ioni{N}{+} and \ioni{S}{+} abundances of about 1~dex in the background-subtracted spectra and 0.5~dex in the observed one.
These variations are related to the stronger dependence of the lines emitted by these ions on the assumed physical conditions. 
There is a further complication in the abundance calculations in proplyds because there may be a density stratification accross the objects.
\cite{rubin89} pointed out that ionic abundances of heavy elements in the ionized gas derived from empirical techniques using CELs can be affected by the presence of density changes along the line of sight.
It is clear that this effect would be even more severe when ionic abundances are calculated making use of emission lines with low critical densities.
Hence, auroral lines should provide more reliable abundances than nebular lines due to their higher critical densities (see Table~\ref{necrit}).
In the particular case of LV2, with a density of \(7.9 \times 10^5~\cmc\), \cite{tsamisetal11} found that the \ioni{O}{2+}, \ioni{N}{+} and \ioni{O}{+} abundances derived from auroral lines were 10\%, 30\% and 50\% higher than those from nebular lines.
Contrary to the results for LV2, we have found an excellent agreement between the \ioni{N}{+}, \ioni{O}{2+} and \ioni{S}{+} abundances derived from nebular and auroral lines.
For the \(\ioni{N}{+}/\ioni{H}{+}\) and \(\ioni{O}{2+}/\ioni{H}{+}\) ratios, the differences of using auroral and nebular lines are between 1\% and 3\% in both background-subtracted and observed spectra, which are equal to the differences obtained in the adjacent background from both kinds of lines for these ions.
For the \ioni{S}{+} abundances, we obtain an abundance difference between auroral and nebular lines of about 5\% (0.02~dex) in the adjacent background.
This difference amounts to 35\% (0.15 dex) in the background-subtracted spectra, but it is of the order of the uncertainty of the \(\ioni{S}{+}/\ioni{H}{+}\) ratio.
The observed spectra show an \(\ioni{S}{+}/\ioni{H}{+}\) abundance from nebular lines that is 0.61~dex higher than that derived from auroral lines.
This is because most of the [\ion{S}{ii}] nebular line emission comes from the nebular background, while auroral lines are emitted from both proplyd and background components.
It is encouraging that we obtain a very good agreement between nebular and auroral lines for the abundances of the ions above, which corroborates the validity of the background subtraction for 177-341 and the physical conditions determined for this proplyd.

 \subsection{Chemical abundances from RLs}  \label{cherl}
For 177-341, we determined the \ioni{He}{+} abundances using the \ion{He}{i} \wav{4471}, \wav{5876}, \wav{6678} lines, which were weighted by 1:3:1 in the computation of the mean values according to their relative line fluxes (see Table~\ref{pro}).
We took the effective recombination coefficients of \cite{storeyhummer95} for \ion{H}{i} and those computed by \cite{porteretal05} with the interpolation formulae provided by \cite{porteretal07} for \ion{He}{i}.
The collisional contribution was estimated from \cite{saweyberrington93} and \cite{kingdonferland95}, and the optical depth in the triplet lines was derived from the computations of \cite{benjaminetal02}.
The adopted physical conditions used to compute the abundances were from the diagnostic diagrams (see Table~\ref{pro}) in both background-subtracted and observed spectra.
For the background emission, we adopted an average value of low ionization zone, \te([\ion{N}{ii}]), and high ionization zone, \te([\ion{O}{iii}]), weighted by the relative ionic fractions \ioni{O}{+}/O and \ioni{O}{2+}/O, respectively.
The results of Table~\ref{pro} show a very similar \ioni{He}{+}/\ioni{H}{+} ratio for the different dust opacity cases.

The good signal-to-noise ratio of our spectral data allows us to detect and measure pure RLs of \ion{O}{ii} and \ion{C}{ii} in most spaxels of the field and, therefore, also in the extracted spaxel of the proplyd 177-341.
In Fig.~\ref{f2} we show a section of the 177-341 spectrum around the lines of Multiplet~1 of \ion{O}{ii}.
RLs of heavy-elements present the advantage that their line flux ratio with respect to an \ion{H}{i} line depends very weakly on the physical conditions, and thus their measurements, and consequently the abundances derived, are much less affected by temperature fluctuations \citep[see][]{peimbert67} and, in our case, the presence of density inhomogeneities.

   \begin{figure*}
   \centering
   \includegraphics[scale=0.95]{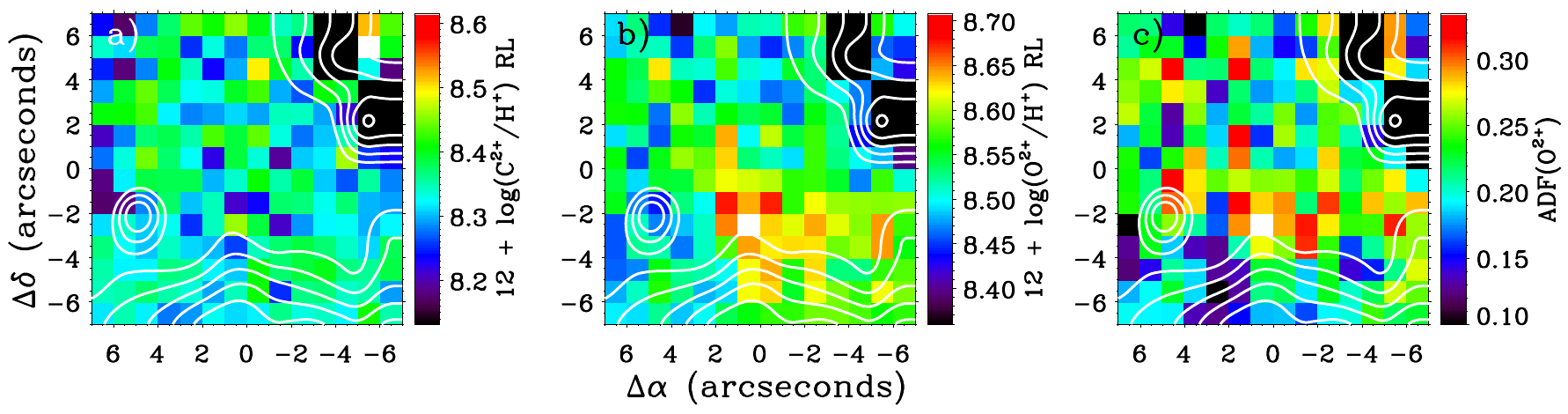} 
   \caption{Spatial distribution maps of \ioni{C}{2+}/\ioni{H}{+} (a) and \ioni{O}{2+}/\ioni{H}{+} (b) abundances derived from RLs and the map of \adfo\ (c). Abundances are presented in units of \(12+ \log(\ioni{X}{+i}/\ioni{H}{+})\). \halpha{} contours are overplotted in all maps. RLs are not detected in the black areas at the upper right corner of the field (positions of proplyds 170-334 and 170-337).}
   \label{f8}
  \end{figure*}

The spectral resolution of our data allows to measure the blend of the brightest lines of Multiplet~1 of \ion{O}{ii}: \ion{O}{ii} \wav{4649}, \wav{4651}.
Then, we used the empirical relations~(1) and~(4) obtained by \cite{apeimbertpeimbert05} to estimate the expected flux of all \ion{O}{ii} M1 multiplet lines with respect to \hbeta{} and determine the \ioni{O}{2+} abundances.
Those expressions also correct for the NLTE effects in the relative flux of each line of the multiplet, though these effects are expected to be rather small in the Orion Nebula because of its relatively high density, and probably negligible in denser proplyds.
Combining the previous empirical relations, we can calculate the \ioni{O}{2+}$/$\ioni{H}{+} ratio from RLs as:
  \begin{equation}
    \frac{\ioni{O}{2+}}{\ioni{H}{+}} = \frac{\lambda_{\mathrm{M1}}}{4861} \times %
                              \frac{\alpha_{\mathrm{eff}}(\hbeta)}{\alpha_{\mathrm{eff}}(\mathrm{M1})} \times %
                              \frac{I(\mathrm{\ion{O}{ii}\,M1})}{I(\hbeta)} ,
  \end{equation}    
where \(\alpha_{\mathrm{eff}}(\hbeta)\) and \(\alpha_{\mathrm{eff}}(\mathrm{M1})\) are the effective recombination coefficients for \hbeta{} and for the \ion{O}{ii} M1 multiplet, respectively, and \(\lambda_{\mathrm{M1}} = 4651.5~\AA\) is the representative mean wavelength of the whole multiplet.
We can also determine the \(\ioni{C}{2+} / \ioni{H}{+}\) ratio from the flux of \ion{C}{ii} \wav{4267} RL using a similar relation.
According to the two-zone scheme defined in \S\ref{cheab}, \te([\ion{O}{iii}]) was adopted to calculate the abundances of the two ions, \ioni{O}{2+} and \ioni{C}{2+}.
We adopted the \ion{O}{ii} effective recombination coefficients from \cite{storey94}, who carried out the calculations in the LS coupling scheme, and the \ion{C}{ii} effective recombination coefficients from \cite{daveyetal00}.
The typical uncertainties amount to 0.11~dex for both \(\ioni{C}{2+}/\ioni{H}{+}\) and \(\ioni{O}{2+}/\ioni{H}{+}\) ratios. 
  
The spatial distribution of the \ioni{C}{2+} and \ioni{O}{2+} abundances from RLs are presented in Fig.~\ref{f8}, and the values obtained from the analysis of the spaxel encompassing 177-341 are shown in Table~\ref{pro}.
It should be noted that we have not detected RLs in the spaxels at and around the proplyds 170-337 and 170-334 (the black areas of Fig.~\ref{f8}).
In Fig.~\ref{f8}(a) the \ioni{C}{2+}/\ioni{H}{+} ratio map shows an almost featureless distribution reaching its highest value near the proplyd 170-334.
However, this may be an artifact because at those spaxels  the \ion{C}{ii} \wav{4267} line is affected by emission features of the neighboring proplyd.
We found that the \ioni{O}{2+}/\ioni{H}{+} ratio map obtained from RLs (Fig.~\ref{f8}b) has similar behaviour to that derived from CELs (Fig.~\ref{f7}b), except for the spaxels \coords(2:0,-3:0) that correspond to the position of the bowshock in front of 177-341, where the \ioni{O}{2+} abundance from RLs shows an enhancement that is not seen in the other maps (see \S\ref{adfo2}).
This fact is not due to a low signal-to-noise ratio of the \ion{O}{ii} RLs because they are rather bright at this zone.
Finally, it is interesting to compare the results of the proplyd 177-341 on the maps with those derived from the observed spectra and included in Table~\ref{pro}.
On the maps and for the same spaxel of 177-341 (see \S\ref{extrac}), we obtain  \ioni{C}{2+} and \ioni{O}{2+} abundances of \(8.32 \pm 0.11\) and \(8.42 \pm 0.11\), respectively, which are very similar to the values derived from the observed spectra case of Table~\ref{pro}, \(8.33 \pm 0.11\) and \(8.40\pm0.11\), respectively.
This is as expected because of the weak dependence on physical conditions of the abundances derived from RLs.
Moreover, due to this weak dependence, the values obtained for the adjacent background emission (\(12 + \log(\ioni{C}{2+}/\ioni{H}{+}) =8.35\pm0.07\) and \(12 + \log(\ioni{O}{2+}/\ioni{H}{+}) = 8.46\pm 0.11\)) and those derived from the background-subtracted spectra (\(12+\log(\ioni{C}{2+}/\ioni{H}{+})=8.25\pm0.12\) and \(12 +\log(\ioni{O}{2+}/\ioni{H}{+})=8.37\pm0.14\)) are very similar considering the relatively low signal-to-noise ratio of these RLs.
Therefore, the different abundance determinations from RLs, including also those of \ioni{He}{+}, remain rather constant and unaffected by the dramatic density contrast that we observe in Table~\ref{pro}.

 \subsection{Abundance discrepancy factor of \ioni{O}{2+}}  \label{adfo2} 
By comparison of the \ioni{O}{2+}/\ioni{H}{+} abundances determined from RLs and CELs we have computed and mapped the AD factor of \ioni{O}{2+}, \adfo, adopting the following definition for this quantity:
\begin{equation}
  \adfo = \log\left(\frac{\ioni{O}{2+}}{\ioni{H}{+}}\right)_{\mathrm{RL}} 
  \kern-0.5em - \log\left(\frac{\ioni{O}{2+}}{\ioni{H}{+}}\right)_{\mathrm{CEL}}.
\end{equation}    
In Fig.~\ref{f8}(c), we can see that the \adfo\ map presents values between 0.10 and 0.35~dex.
We do not compute the \adfo\ at 170-337 and 170-334 and their immediate surroundings because the \ion{O}{ii} RLs are not detected in those areas.
Excluding the proplyd positions, we obtain an average \adfo\ of about 0.23~dex with a standard deviation of 0.06~dex for the whole map.
This value is similar to previous determinations in other parts of the nebula \citep{estebanetal04, mesadelgadoetal08, mesadelgadoetal09a, mesadelgadoetal11}.
As mentioned in the previous section, the spaxels at the position of the bowshock related to 177-341 show an increase in the flux of \ion{O}{ii} RLs but not in the [\ion{O}{iii}] CELs.
This produces an \adfo\ of about 0.29~dex, while its immediate surroundings show values about 0.22~dex.
Deeper and more detailed spectroscopic observations are needed in order to confirm and explore this anomaly. 

From Table~\ref{pro} we can see that the \adfo\ of the observed spectrum of 177-341 amounts to \(0.04 \pm 0.11\)~dex, while in the \adfo\ map the value of that precise spaxel is about 0.30~dex.
This dramatic difference is related to the \te\ and \nel\ bias affecting the CELs when nebular contamination is not taken into account.
The reasoning is the following: a) we use the densities of the \nel([\ion{S}{ii}]) map, which are too low for the proplyds as discussed in \S\ref{density}, for the calculations; b) the \te([\ion{O}{iii}]) map derived using those densities produces artificially higher temperatures at the proplyds; and c) the \ioni{O}{2+} abundances calculated from CELs in this way are lower than expected.
In Table~\ref{pro} we can also see that the \adfo\ of the background-subtracted spectra of 177-341 is \(-0.06\pm0.15\)~dex.
Comparing the \ioni{O}{2+} abundances determined for the different cases in Table~\ref{pro}, it can be seen that the change of the \adfo\ from one case to the other is mainly due to variations of the \ioni{O}{2+}/\ioni{H}{+} ratio derived from CELs.
Those variations are larger than the quoted errors, in contrast to what happens for the abundances determined from RLs, which can be considered basically the same within the errors in the background-subtracted spectra and in the observed one. 
A similar result is found for the other ionic abundances obtained from RLs: \ioni{He}{+} and \ioni{C}{2+}.
Therefore, the variations of \ioni{O}{2+}/\ioni{H}{+} ratio derived from CELs may be produced by the temperature structure changes related to the presence of high-density ionized gas (and perhaps strong density stratification) in the proplyd.
In contrast, abundances derived from RLs remain unaffected due to their weaker dependence on the physical conditions.
Similar results were also found in the proplyd LV2 analysed by \cite{tsamisetal11} and clearly show that the presence of high-density ionized gas is playing a role in the AD problem.

Based on an IFU analysis of LV2 and the surrounding nebula, and taking into account theoretical considerations \citep{rubin89, viegasclegg94}, \cite{tsamisetal11} concluded that an important contribution to the AD problem in \ion{H}{ii} regions is the presence of dense regions (e.g., proplyds, globules or filaments).
If these results can be generalized, the existence of a population of high-density, small and semi-ionized clumps/filaments, mixed with the nebular background gas of the Orion Nebula and randomly located along the line of sight, may be affecting the classic [\ion{O}{iii}] \wav{4363}/\wav{4959} forbidden line temperature ratio within the observation aperture and, therefore, the gaseous abundances derived from CELs.
In that case, CELs would yield lower limits to the mean Orion abundances, which would be more reliably determined from the faint metallic RLs.
This last possibility is supported by \cite{simondiazstasinska11}, who conclude that the stellar O abundances of the Orion OB Association are more consistent with gaseous ones determined from RLs (after correcting for the oxygen depletion onto dust grains).
Further to be addressed is the question of how any such density inhomogeneities relate to the classic temperature fluctuation paradigm \citep{peimbert67}, which has been invoked as a possible solution to the AD problem.
Follow-up studies on the link between the \adfo\ and temperature fluctuation parameter, \tf, determinations are needed.

Considering the results of LV2 \citep{tsamisetal11} and 177-341 (this work), we can conclude that these unknown clumps, similar in properties to the proplyds, need not be strong metallic RL emitters relative to the background emission of the Orion Nebula.
This is supported by the fact that both studied proplyds show line flux ratios with respect to \hbeta{} of the RLs \ion{He}{i}, \ion{O}{ii} and \ion{C}{ii} that are similar to those of the nebular background (see Table~\ref{pro} for 177-341).
This line of reasoning is different from that of previous scenarios in which density inhomogeneities in \ion{H}{ii} regions were assumed to be hydrogen-deficient with respect to the host nebula and strong RL emitters on account of their low electron temperatures \citep{tsamispequignot05, stasinskaetal07}.
Another important issue is the size of the posited clumps and their spatial distribution, which would determine the total amount of ionizing radiation that they receive and the spectrum that they emit.
In this sense, previous studies performed in the Orion Nebula \citep[e.g.][]{rubinetal03, mesadelgadoetal08} have not found any evidence for clumps, other than the known proplyds.
The work carried out by \cite{rubinetal03} based on long-slit spectroscopy obtained with the \textit{HST} attained a spatial resolution of 0\farcs5, while \cite{mesadelgadoetal08} studied the spatial distribution of physical conditions and abundances at 1\farcs2 from ground-based observations.
Neither study detected significant small-scale density variations in the nebular background.
On the one hand, in \cite{mesadelgadoetal08} the measurements were made of \nel([\ion{S}{ii}]), which can be affected by collisional de-excitation in the presence of high-density gas.
On the other hand, \cite{rubinetal03} used a low density limit for the temperature determinations.
So, it is clear that deeper spectroscopic studies using the \textit{HST} at the highest spatial resolution (e.g., with WFC3 in the UV/far-UV), along with the use of appropriate density indicators are needed to explore the astrophysics of small scale ``clumpiness" in the Orion Nebula.
Finally, it should be noted that if these kinds of clumps are real, or simply there are important density variations in the nebular background, they would severely modify the CEL diagnostics. The existence of dense clumps within more distant Galactic and extragalactic \hii\ regions can complicate the interpretation of emission line diagnostics with important ramifications for chemical evolution studies.

 \section{Dust effects on nebular properties of the proplyd 177-341} \label{dust}
Dust particles in the inner dense neutral zones of the proplyds can absorb and scatter the emission from the surrounding nebula, 
which complicates the background subtraction process and the determination of the intrinsic proplyd emission.
However, the dust properties of the proplyds are not well known.
The only existing study is that of  \cite{rostetal08}, who attempted to model near-infrared polarization mapping of 177-341, but found that simple models do not reproduce the observations.
We therefore adopt a simplified empirical procedure,  similar to that presented in Figure~3a of \cite{henneyodell99}. 
The total dereddened flux for a given wavelength, $I_\lambda\total$, measured at a proplyd position can be expressed schematically as:
\begin{equation}
  I_\lambda\total = I_\lambda\bg e^{-\tau_\lambda\pe} + I_\lambda\proplyd, 
\end{equation}
where $I_\lambda\bg$ represents the nebular background emission, which is attenuated by an effective opacity proplyd optical depth, $\tau_\lambda\pe$, and $I_\lambda\proplyd$ is the intrinsic proplyd emission.

The effective optical depth will be significantly smaller than the true absorption optical depth of the proplyd due to three effects.  
First, some portion of the nebular emission may arise from in front of the proplyd and so will not be absorbed. 
Second, the neutral core of the proplyd is small compared with the angular resolution of the observations, which produces ``beam dilution'' of the absorption. 
Third, scattering of nebular light by dust in the proplyd will partially ``fill in'' the absorption. 
This last effect will depend on the geometrical relation between the proplyd and the bulk of the emitting gas in the nebula (as well as on the dust properties).
For 177-341, the scattered contribution might be relatively high because we clearly see the proplyd projected against a particularly faint part of the nebula, whereas there are much brighter parts off to the side.
In principle, we could constrain this by analysing the scattered continuum, but it requires further investigation, which is beyond the scope of this article.

In the absence of clear observational constraints, we adopt a standard reddening law, \(f(\lambda, R_V\proplyd)\), parametrized by the total-to-selective extinction, $R_V\proplyd$ \citep{cardellietal89}, which will be a function of the grain optical properties and the relative importance of scattering. 
The proplyd opacity at any wavelength \(\lambda\) is then related to the \hbeta{} extinction coefficient by 
\begin{equation}
  \tau_\lambda\pe = \frac{\chb\proplyd}{\log e}\left(1 + f(\lambda, R_V\proplyd)\right)
\end{equation} 
and the absolute dereddened flux of a proplyd is given by
\begin{equation}
  I_\lambda\proplyd = I_\lambda\total - I_\lambda\bg 10^{-\chb\proplyd\, \left( 1+ f(\lambda,R_V\proplyd)\right)}.
\end{equation} 
Assuming that the foregroud extinction, \chb\los, is the same for the proplyd and adjacent nebular emission, then the raw observed fluxes (\(F_\lambda = I_\lambda e^{-\tau_\lambda\los}\)) follow a parallel relation: 
\begin{equation}
  F_\lambda\proplyd = F_\lambda\total - F_\lambda\bg 10^{-\chb\proplyd \, \left( 1+ f(\lambda,R_V\proplyd)\right)}. 
\end{equation}
This assumption is justified for the proplyd 177-341 since it can be seen from Table~\ref{pro} that the \chb\los{} coefficients derived from the background-subtracted, observed, and background spectra are identical within their respective uncertainties.

It is also interesting to note that the cases presented in \S\ref{extrac} --and analysed along this paper for the different proplyds-- can be easily obtained from Eq.~6.
We can see that the transparent case, \(\chb\proplyd =0 \), is just a simple subtraction of the nebular spectra from the total observed spectra, where both nebular and proplyd emissions are mixed.
On the contrary, in the opaque case, the proplyd spectra directly corresponds to the observed one.
An interesting point of these two cases is that neither of them depend on the dust properties of the proplyd.

These relations allow us to go beyond the approximations used in the previous sections, in which the proplyd was assumed to be either fully opaque (\(\chb\proplyd \gg1 \)) or fully transparent (\(\chb\proplyd =0 \)), and to explore the effects of partial transparency on the derived physical conditions and chemical abundances of 177-341. 
The final results are presented in Table~\ref{pro}.  
As well as for the transparent and opaque cases, we show diagnostics derived assuming \(\chb\proplyd = 0.05\) and 0.1~dex, which are similar to the values determined by \cite{henneyodell99} for this proplyd (0.04 to 0.08~dex).
This pair of values is combined with three different values of $R_V\proplyd = 3.1$, 5.5 and 7, using the $f(\lambda)$ reddening law of \cite{blagraveetal07}.
Once we have estimated the proplyd fluxes, $F_\lambda\proplyd$, for the different cases, we proceed with the nebular analysis calculations: \chb\los, physical conditions and chemical abundances.
As we see in Table~\ref{pro}, the standard ISM value for $R_V\proplyd$ of 3.1, produces \chb\los{} coefficients that are similar between the proplyd and background emissions, while higher $R_V\proplyd$ would imply lower \chb\los for the proplyd.
This indicates that the dust properties in the proplyd 177-341 might be better represented by $R_V\proplyd=3.1$, though the uncertainties are too large to reach a solid conclusion.

With respect to the physical conditions, we can note that the \nel\ determinations are the most affected by the proplyd extinction.
For $R_V\proplyd=3.1$, the adopted density falls from \(3.8 \times 10^5\) to \(2.4 \times 10^5~\cmc\) when \(\chb\proplyd  \) increases from 0 to 0.1 dex.
A similar contrast is found for the other two values of $R_V\proplyd$.
However, the electron temperatures do not show large differences.
Indeed, they slightly decrease for higher $R_V\proplyd$ ratios.
For the same \(\chb\proplyd\) cases and $R_V\proplyd=3.1$, \te([\ion{N}{ii}]) and \te([\ion{O}{iii}]) show variations of 200 and 130~K, respectively.
The derived chemical abundances are in turn affected by the changes in the physical conditions.
The ionic abundances of \ioni{N}{+}, \ioni{Ne}{2+}, \ioni{S}{+} and \ioni{S}{2+} are the most affected by \(\chb\proplyd \) variations because their abundance determinations have a stronger density dependence than the other ions.
The largest differences are found for \ioni{S}{+}/\ioni{H}{+} and \ioni{N}{+}/\ioni{H}{+} ratios: 0.24 and 0.19~dex, respectively when \chb\proplyd{} varies from 0 to 0.1, which is larger than the quoted uncertainties.
Due to the much weaker dependence on the assumed physical conditions, ionic abundances of \ioni{He}{+}, \ioni{C}{2+} and \ioni{O}{2+} determined from RLs remain unaffected.
 
 \section{A physical model of 177-341} \label{model}
As an alternative to the purely empirical analysis presented in the previous sections, a different approach to analysing the emission spectrum of the proplyds is through the construction of physical models that combine a priori simulations of radiative transfer, hydrodynamics, and atomic physics in order to predict the density, temperature, and ionization structure of the the photoevaporation flow through the proplyd ionization front. 
Such models have previously been applied to the ensemble properties of large numbers of proplyds \citep*{johnstoneetal98, henneyarthur98} and in detail to individual objects such as 177-341 \citep{henneyodell99}, LV2 \citep{henneyetal02}, and LV1 \citep{grahametal02}. 
We have improved on these models in two significant ways. 
First, whereas published models have considered only emission from regions where hydrogen is fully ionized and the flow is supersonic, we now use a detailed analytic model of gas acceleration in the ionization front \citep{henneyetal05} to extend the treatment to cover partially ionized emission zones where the gas moves subsonically. 
Second, whereas published models used ad hoc fitting functions to the emissivity structure, specifically tailored to only the brightest emission lines, we now use the plasma microphysics code Cloudy \citep{ferlandetal98} to self-consistently calculate the full physical structure and emission spectrum of the proplyd flow.

\begin{table}
  \centering
  \caption{Input parameters for example physical model of 177-341} 
  \begin{tabular}{@{\,}ll@{\,}}\hline
    Stellar spectrum:& 
    \(T_* = \SI{39000}{K}\)\\
    \citep{simondiazetal06} & \(\log g = 4.1\)\\
    & \(L_* = \num{2.04e5}\,L_\odot\)\\
    Ionizing flux at proplyd:& 
    \(\Phi_{\mathrm{H}} = \SI{1.58e13}{cm^{-2}\ s^{-1}}\)
    \\
    Ionization front radius:& 
    \(r_0 = \SI{1.91e15}{cm}\)
    \\
    Gas-phase abundances: & 
    He 10.98, C 8.41, N 7.85, O 8.30, \\
    (\(12 + \log z/\mathrm{H}\)) & Ne 7.56, S 6.98, Ar 6.26, Fe 5.65
    \\
    Dust composition: & Standard Orion \citep{baldwinetal91}\\
    \hline
  \end{tabular}
  \label{tab:model:pars}
\end{table}

\begin{figure*}
  \centering
  \begin{tabular*}{\linewidth}{@{\extracolsep{\fill}} ll}
    (\textit{a}) & (\textit{b}) \\
    \includegraphics[width=0.45\linewidth]{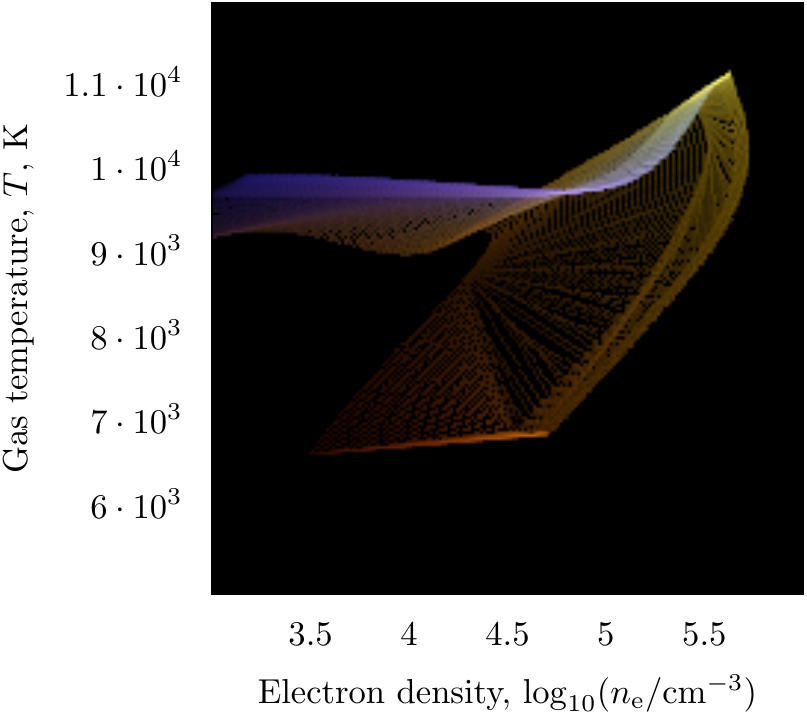} &
    \includegraphics[width=0.465\linewidth]{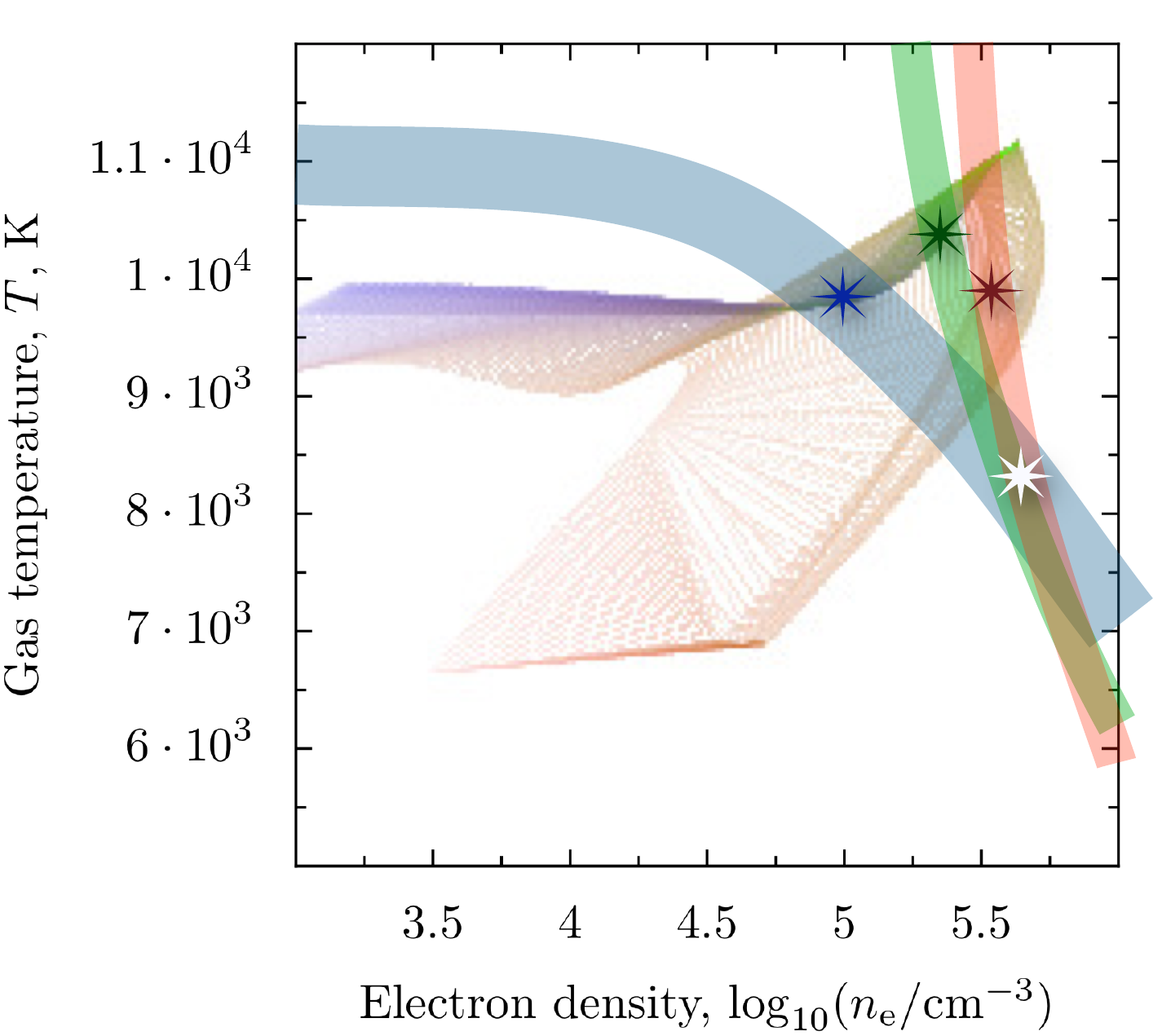}
  \end{tabular*}
  \caption[este es el caption]{Emission structure in the \nel--\te\ plane for the example physical model of 177-341. (a) Positive-colored image of the three emission lines: [\ion{S}{ii}] \wav{6731} (red), [\ion{N}{ii}] \wav{6583} (green), and [\ion{O}{iii}] \wav{5007} (blue), with brightness proportional to the fraction of the model luminosity in each line. (b) Negative-colored image of the same data overlaid with the observational diagnostic curves from Fig.~\ref{f61} ([\ion{S}{ii}] in red, [\ion{N}{ii}] in green, [\ion{O}{iii}] in blue).}
  \label{fig:model:nT}
\end{figure*}
\begin{figure*}
  \centering
  \includegraphics[scale=0.7]{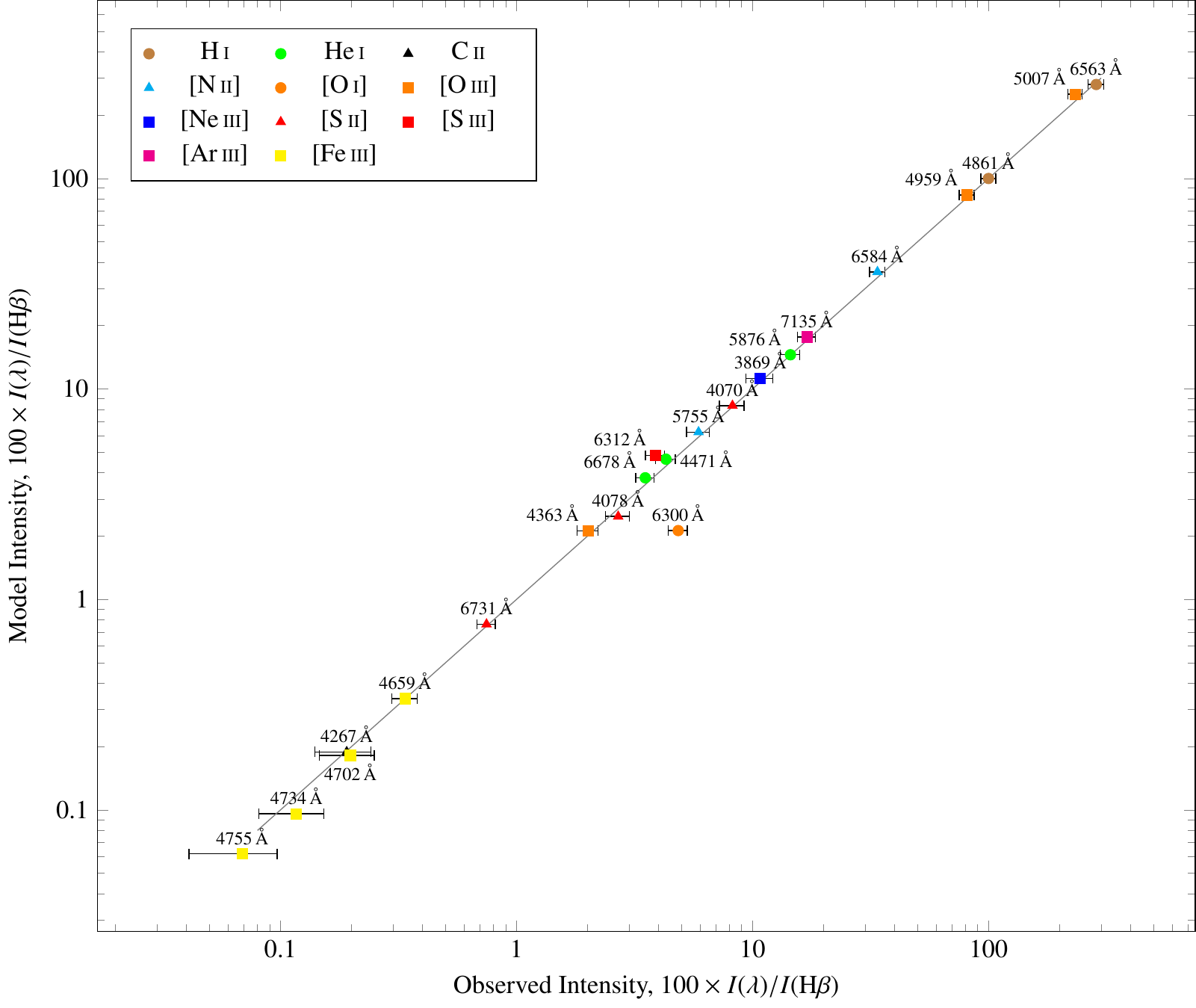}
  \caption{Comparison of the model and observed line ratios for the proplyd 177-341. The model parameters are given in Table~\ref{tab:model:pars} and the emission structure of the model is presented in Fig.~\ref{fig:model:nT}. Observed intensities correspond to the background-subtracted values (\(\chb\proplyd =0 \)) given in column~4 of Table~2.}
  \label{fig:model}
\end{figure*}

The input parameters for the model (values given in Table~\ref{tab:model:pars}) are the radius of the ionization front at the proplyd cusp (assumed hemispherical), the intensity and spectral shape of the illuminating stellar radiation, and the composition (gas-phase elemental mix and dust grain populations) of the proplyd material. 
For most of these parameters, we have taken values from the literature, whereas the gas-phase abundances and incident ionizing flux have been adjusted slightly in an attempt to reproduce the observed emission line intensities. 
The stellar spectrum is that determined for the ionizing star \thC{} by spectroscopic analysis \citep{simondiazetal06}. 
The ionization front radius is the value determined by fitting to \textit{HST} emission line images \citep{henneyodell99}, while the adopted ionizing flux at the proplyd position corresponds to a physical separation of the proplyd from the ionizing star of \(2.13\times10^{17}\)~cm if there is no intervening absorption. 
Given the observed angular separation of 25\farcs84 \citep{ballyetal98}, and assuming a distance to the Orion Nebula of 440~pc \citep{odellhenney08}, the projected separation is \(1.70\times10^{17}\)~cm, implying an inclination angle of $\simeq 55^\circ$. 
The diffuse radiation field and the proplyd tail are ignored in this model, since the observational aperture (see \S\ref{extrac}) only covers the head of the proplyd. 

The basic structure of the model is of a slow ionized wind that expands spherically away from the ionization front at the bright proplyd cusp \citep[see figure~1 of][]{henneyarthur98}. 
The outer layers of this externally ionized wind have the highest ionization parameter because they are relatively low density (\(< 1000~\cmc\)) and see the full ionizing flux from the Trapezium stars. 
The inner layers of the wind are denser and see a reduced flux because of absorption in the overlying layers. 
This gives a lower ionization parameter and also a higher temperature due to radiation hardening and collisional quenching of coolant lines. 
The temperature reaches a maximum value of \(\sim 11000\)~K just outside the ionization front, while the electron density peaks inside the front itself. 
For the innermost layers, hydrogen becomes increasingly neutral and both the temperature and electron density drop rapidly. 
Thus, each radius within the proplyd has its own combination of density, temperature, and ionization state. 
It is the correlations between these quantities (as illustrated in Fig.~\ref{fig:model:nT}) that
determine the optical emission line spectrum of the proplyd (Fig.~\ref{fig:model}).

Fig.~\ref{fig:model:nT}(a) shows the emission structure of [\ion{S}{ii}] \wav{6731} (red), [\ion{N}{ii}] \wav{6583} (green), and [\ion{O}{iii}] \wav{5007} (blue). 
The low-density outer zones of the model are highly ionized, emitting principally in [\ion{O}{iii}], as seen in the blue horizontal branch at \(\sim 9700\)~K in the upper left part of the figure. 
As one approaches the proplyd ionization front (direction of increasing density), the [\ion{N}{ii}] and then [\ion{S}{ii}] emissions become relatively stronger (color changes to yellow) and the temperature increases, reaching a maximum of \(\sim 11000\)~K. In the ionization front itself, where the gas is partially neutral, the [\ion{S}{ii}] emission comes to dominate over [\ion{N}{ii}], giving rise to the orange-red branch that curves down and to the left. Fig.~\ref{fig:model:nT}(b) shows a negative representation of the same data as in Fig.~\ref{fig:model:nT}(a) overlaid with the observational diagnostic curves from Fig.~\ref{f61}(a) ([\ion{S}{ii}] in red, [\ion{N}{ii}] in green, [\ion{O}{iii}] in blue). 
The white star shows the solution for (\nel, \te) obtained in \S\ref{phycon} under the assumption that the emission in all three ions is co-extensive at a single density and temperature. 
This solution lies well outside the locus of densities and temperatures seen in the model, even though the model does reproduce the observed line ratios (see Fig.~\ref{fig:model}). 
Instead, the model satisfies the constraints of the three diagnostic curves separately in three different regions, which are indicated schematically by the colored stars. 
Note, however that even within the emission region of each ion there is considerable variation in the physical conditions. 
Independent evidence that the higher proplyd temperatures deduced from the physical model may be correct comes from the temperature map of \cite{garciadiazetal08}, who found temperature peaks of \num{10000} to \SI{12000}{K} at the positions of several proplyds, including 177-341 (see their figure 21). 
Those authors used a linewidth method instead of the traditional line ratio methods, and so their results should be insensitive to uncertainties in the electron density.

Fig.~\ref{fig:model} shows that the model generally reproduces very well the observed relative line intensities for the intrinsic spectra of 177-341 with the notable exception of the [\ion{O}{i}] \wav{6300}  line, which has a predicted flux of only half the observed value. 
This discrepancy may be explained if a significant fraction of the observed [\ion{O}{i}] emission comes from dissociation of OH at the surface of the proplyd's circumstellar disc \citep{storzerhollenbach98}, which is not included in our model. 
Another factor that may increase such discrepancy is that the [\ion{O}{i}] emission in the observed and background spectra has not been corrected from telluric emission contamination, though the subtraction process should minimize it in the intrinsic spectra of the proplyd. 
The only other disagreement is with the [\ion{S}{iii}] \wav{6312} line, which is roughly 25\% too high in the model (roughly 1.5 times the observational uncertainty). 
This may be an indication that the details of the stellar spectrum around 23~eV are not correctly modeled, although the disagreement is only marginal. 

Given that the stellar spectrum is very well constrained, the only way to vary the model temperature by a significant amount is by changing the gas-phase metal abundances and thus the gas cooling rate. 
In particular, the temperature in the highly ionized outer regions of the proplyd is very sensitive to the oxygen abundance since the [\ion{O}{iii}] \wav{5007} line is the major coolant there. 
In order to reproduce the observed line intensities (particularly the [\ion{O}{iii}] \wav{4363}/\wav{5007} ratio), we find it necessary to reduce the oxygen abundance to less than 50\% of the value that has been measured for the surrounding nebula (e.g., \citealp{estebanetal04}), which leads to a temperature of around 9500~K in the [\ion{O}{iii}] emission zone. 
Using the standard Orion gas-phase abundances gives temperatures of about 8500~K for the same zone and predicts [\ion{O}{iii}]/\hbeta{} ratios that are considerably higher than observed for the nebular [\ion{O}{iii}] \wav{4959}, \wav{5007} lines, but lower than the observed for [\ion{O}{iii}] \wav{4363} auroral line. 
Using the even higher abundances derived by \citet{tsamisetal11} for the proplyd LV2 implies temperatures as low as 7700~K and exacerbates the disagreement with the observations still further. 
For comparison, both varying the stellar effective temperature by $\pm1000$~K (the uncertainty in the analysis of \citealp{simondiazetal06}) or varying the ionizing flux by a factor of two (corresponding to the uncertainty in the inclination angle of the proplyd) only change the gas temperature by $<200$~K, while varying the stellar atmosphere model between WMBasic \citep{pauldrachetal01} and TLUSTY \citep{lanzhubeny03} produces a slightly larger change of $\simeq 400$~K.

Despite the promising success of the model in reproducing the observed line ratios of the intrinsic spectra of 177-341, further work is necessary before a definitive statement can be made about the gas-phase abundances in the proplyd. 
The lack of observations of [\ion{O}{ii}] lines, together with the uncertain nature of the [\ion{O}{i}] emission (see above) means that the oxygen abundance hinges on observations of a single ion stage. 
It is vital to test the model against other proplyds, particularly those such as LV2 \citep{tsamisetal11, tsamiswalsh11} where the [\ion{O}{ii}] \wav{3726}, \wav{3729} doublet is observed. 
At the same time, the models need to be simultaneously constrained not just by the line ratios, but also by sub-arcsecond \textit{HST} imaging \citep{ballyetal98, odell98} and velocity profiles from high-resolution spectroscopy \citep{henneyodell99, shupingetal03}. 
It is also necessary to examine in greater detail the potential influence of internal extinction within the proplyd. 
This depends on the properties of the dust that is entrained in the photoevaporation flow, which can be constrained by comparison with observations of thermal mid-infrared emission \citep{garciaarredondoetal01, smithetal05, robbertoetal05, shupingetal06}.

\section{Conclusions} \label{conclu}
In this paper we used integral field spectroscopy in the spectral range 3500--7200~\AA\ to explore the effect of the presence of protoplanetary discs (proplyds) on the controversial abundance discrepancy (AD) problem. 
Our field is located to the south of the Trapezium Cluster and contains different morphological structures. 
Among them, the most prominent structures are the proplyds 177-341, 170-337 and 170-334 and the bowshock in front of 177-341 and the high-velocity jet HH~514 emerging from 170-337.  
 
We analyse spatial distribution maps of several nebular properties: emission line fluxes, extinction coefficient, electron densities and temperatures, chemical abundances from collisionally excited lines (CELs) and recombination lines (RLs), and the AD factor of \ioni{O}{2+}, \adfo. 
We found that collisional de-excitation due to high densities is the main mechanism to explain the behaviour observed in the spatial distributions of the nebular properties at the position of the proplyds. 
Analysing the collisional de-excitation effects, we estimate that proplyds 177-341 and 170-334 should have a density between \(9\times10^4\) and \(7\times10^5~\cmc\), the critical densities of the departure levels of the [\ion{N}{ii}] and [\ion{O}{iii}] nebular lines, respectively. 
For the proplyd 170-334, we estimate that it should present a density higher than \(7 \times 10^5~\cmc\), the critical density of [\ion{O}{iii}] nebular lines, based on the intense enhancement observed for this proplyd in the spatial distribution of the [\ion{O}{iii}] auroral line. 
 
In this study we also obtain the intrinsic spectra of the proplyds 177-341, 170-337 and 170-334 by subtracting the nebular background emission. 
From a detailed analysis, we derive densities, temperatures and chemical abundances using empirical diagnostics for the three proplyds. 
Due to their smaller sizes, the diagnostics reveals that proplyds 170-337 and 170-334 present some background contamination, so the physical conditions and chemical abundances derived for them may not be correct. 

We detect \ion{O}{ii} and \ion{C}{ii} RLs in the spectrum of the proplyd 177-341. 
The density determined in its background-subtracted spectrum is about \(3.8 \times 10^5~\cmc\),
which is much higher than the density of the nebular background of the Orion Nebula, \(3800~\cmc\). 
We determine electron temperatures from [\ion{N}{ii}] and [\ion{O}{iii}] line ratios finding that the striking spikes observed in the spatial temperature distribution maps 
disappear once the correct density value for the proplyds is used.. 
We obtain the chemical abundances of an extensive set of ions from CELs: \ioni{N}{+}, \ioni{O}{2+}, \ioni{Ne}{2+}, \ioni{S}{+}, \ioni{S}{2+}, \ioni{Ar}{2+} and \ioni{Fe}{2+}. 
We also determine ionic abundances from RLs for \ioni{He}{+}, \ioni{O}{2+} and \ioni{C}{2+} and observe that ionic abundances from CELs show important differences when comparing the observed spectrum and the intrinsic one of 177-341, while those derived from RLs remains similar. 
We found that ionic abundances of \ioni{N}{+}, \ioni{O}{2+} and \ioni{S}{+} derived from nebular and auroral lines are very consistent with one another when using the correct high densities. 

We investigate the effects on our results of internal extinction by dust in the proplyd's neutral core. 
We find that the resultant uncertainty in the background subtraction process may have a significant impact on the derived density and ionic abundances (such as \ioni{S}{+}/\ioni{H}{+} and \ioni{N}{+}/\ioni{H}{+}) that are sensitive to this. 

As an alternative to the empirical analysis we construct a physical model for the proplyd 177-341, obtaining the ionization structure of the proplyd as well as its temperature and density. 
We find an excellent agreement between the observed line ratios of the intrinsic spectra and the predicted ones from the model, even though the predicted physical conditions for the well-resolved \ioni{S}{+}, \ioni{N}{+} and \ioni{O}{2+} zones differ from those obtained in the one-zone scheme applied in the empirical analysis. 
Regarding the final gas-phase abundance in the proplyd, further work is needed to definitively determine them.
 
Finally, comparing the \ioni{O}{2+}/\ioni{H}{+} ratio derived from CELs and RLs in the proplyd 177-341, we find that the \adfo\ tends to 0 in the observed spectra and in the intrinsic spectra of this proplyd. 
A similar result was previously found in the analysis of the proplyd LV2 performed by \cite{tsamisetal11}. 
Though our determinations show relatively high uncertainties, we conclude that high-density gas (in the form of proplyds, globules or unseen clumps/filaments acting at small spatial scales) is playing a major role in the AD problem. 
Based on the results on 177-341 and LV2, we discuss the AD problem in the context of density inhomogeneities finding that the proposal of \cite{viegasclegg94} seems to explain the behaviour observed in the proplyds. 
In a scenario where small, high-density and semi-ionized clumps/filaments are mixed with diffuse gas in the observation aperture, the classical CEL abundance analysis can be affected if collisional de-excitation of certain emission line diagnostics is not accounted for. 
RLs would not be affected and, therefore, they should reliably yield the chemical abundances in the target field. 
Judging by the results of the proplyd IFU analyses thus far, these clumps need not be strongly hydrogen-deficient, unlike those posited in previous scenarios to explain the AD problem in \ion{H}{ii} regions \citep{tsamispequignot05, stasinskaetal07} and planetary nebulae \citep{harringtonfeibelman84, liuetal00, tsamisetal08}.

\section*{Acknowledgments} 
The referee is thanked for a detailed and very helpful report.
We are also grateful for the support offered by C.~Morisset and V.~Luridiana in the use of the new software {\sc pyneb}. 
WJH and NFF acknowledge financial support from DGAPA-UNAM through project PAPIIT IN102012 and from a postdoctoral fellowship to NFF. 
YGT acknowledges the award of a Marie Curie Intra-European Fellowship within the 7th European Community Framework Programme (grant agreement PIEF-GA-2009-236486). 
This work has been funded by the Spanish Ministerio de Educaci\'on y Ciencia (MEC) and Ministerio de Ciencia e Innovaci\'on (MICINN) under projects AYA2007-63030 and AYA2011-22614.



\label{lastpage}

\end{document}

%% file: tabla_necrit.tex
 {[\ion{N}{ii}]} & 6548, 6584 & 9$\times$10$^4$              & 24 & 24 & 12\\
                                     &          5755 &                10$^7$             & 77 & 61 & 56\\
{[\ion{O}{ii}]} & 3726, 3729 & 10$^3$, 4$\times$10$^3$ & 0 & 17 & 9\\
                                     &  7318, 7329 &  4$\times$10$^6$            & -- & -- & -- \\
{[\ion{O}{iii}]} & 4959, 5007 & 7$\times$10$^5$            & 24 & 22 & 27 \\
                                       &          4363 & 2$\times$10$^7$            & 37 & 38 & 51\\
                   {[\ion{Ne}{iii}]} & 3869 & 10$^7$                                    & 17 & 35 & 27 \\
{[\ion{S}{ii}]} & 6717, 6731 & 2$\times$10$^3$, 4$\times$10$^3$ & 5 & 17 & 4 \\
                                     & 4068, 4076 & 8$\times$10$^5$, 2$\times$10$^6$ & 75 & 54 & 39 \\ 
                    {[\ion{S}{iii}]} & 6312 & 10$^7$                                     & 41 & 32 & 32 \\                    
                    {[\ion{Ar}{iii}]} & 7135 & 5$\times$10$^6$                      & 28 & 30 & 26 \\

%% file: tablalines.tex
3728.00  &    [\ion{O}{ii}]  &         1F  &    	 		     --&    50.6$\pm$3.7 &    85.1$\pm$6.0 &    41.2$\pm$4.5 &    71.1$\pm$7.3 &    84.7$\pm$5.9&    25.4$\pm$3.2 &   69.8$\pm$5.0&    83.8$\pm$6.0\\
3868.75  &  [\ion{Ne}{iii}]  &         1F  &    10.8$\pm$1.4 &    18.6$\pm$1.3 &    21.6$\pm$1.4 &    25.5$\pm$2.7 &    21.6$\pm$2.0 &    20.7$\pm$1.4 &    23.7$\pm$2.8 &   22.2$\pm$1.5&    20.7$\pm$1.4\\
4068.60  &    [\ion{S}{ii}]  &         1F  &    8.23$\pm$0.98&    3.27$\pm$0.26&    1.08$\pm$0.08&    4.25$\pm$0.46&    2.27$\pm$0.18&    1.44$\pm$0.11&    2.95$\pm$0.35&    1.70$\pm$0.13&    1.21$\pm$0.09\\
4069.68  &      \ion{O}{ii}  &         10  &    		&    		     &    		&    			&    		&    			&    	&    		&    	\\
4069.89  &      \ion{O}{ii}  &         10  &    		&    		     &    		&    			&    		&    			&    	&    		&    	\\
4076.35  &    [\ion{S}{ii}]  &         1F  &    2.67$\pm$0.32&    1.15$\pm$0.09&    0.46$\pm$0.05&    1.61$\pm$0.17&    0.92$\pm$0.07&    0.55$\pm$0.06&    0.57$\pm$0.09&    0.60$\pm$0.07&    0.49$\pm$0.06\\
4101.74  &       \ion{H}{i}  &  H$\delta$  &    21.9$\pm$2.3 &    24.8$\pm$1.5 &    25.7$\pm$1.5 &    25.8$\pm$2.4 &    25.9$\pm$1.6 &    25.8$\pm$1.5 &    22.4$\pm$2.3 &    24.9$\pm$1.5 &    25.5$\pm$1.5 \\
4267.15  &      \ion{C}{ii}  &          6  &    0.19$\pm$0.05&    0.23$\pm$0.06&    0.24$\pm$0.04&    	             --&    	          --&    0.21$\pm$0.05&    	            --&    	         --&    0.19$\pm$0.05\\
4340.47  &       \ion{H}{i}  &  H$\gamma$  &    47.9$\pm$4.4 &    48.0$\pm$2.6 &    47.3$\pm$2.5 &    47.1$\pm$4.0 &    46.9$\pm$3.1 &    47.1$\pm$2.5 &    48.0$\pm$4.4 &    47.9$\pm$2.6 &    47.6$\pm$2.6 \\
4363.21  &   [\ion{O}{iii}]  &         2F  &    2.01$\pm$0.21&    1.65$\pm$0.12&    1.40$\pm$0.10&    1.85$\pm$0.18&    1.43$\pm$0.12&    1.45$\pm$0.10&    5.29$\pm$0.54&    2.60$\pm$0.19&    1.41$\pm$0.10\\
4471.47  &      \ion{He}{i}  &         14  &    4.30$\pm$0.42&    4.52$\pm$0.32&    4.59$\pm$0.32&    4.09$\pm$0.38&    4.27$\pm$0.33&    4.34$\pm$0.30&    4.09$\pm$0.39&    4.45$\pm$0.31&    4.46$\pm$0.31\\
4607.13  &  [\ion{Fe}{iii}]  &         3F  &    	     --&    0.02$\pm$0.01&    0.04$\pm$0.02&    	         --&		     --&		 --&  		     --&		     --&		  --\\
4649.13  &      \ion{O}{ii}  &          1  &    0.14$\pm$0.04&    0.15$\pm$0.04&    0.15$\pm$0.04&    	         --&		     --&    0.17$\pm$0.04&		     --&		     --&    0.21$\pm$0.05\\
4650.84  &      \ion{O}{ii}  &          1  &    	       &    		   &    	       &    		   &		       &		   &	    	       &		    &	    \\
4658.10  &  [\ion{Fe}{iii}]  &         3F  &    0.34$\pm$0.04&    0.46$\pm$0.05&    0.52$\pm$0.06&    	         --&	0.84$\pm$0.06&    0.49$\pm$0.05&			--&		     --&    0.49$\pm$0.05\\
4701.62  &  [\ion{Fe}{iii}]  &         3F  &    0.20$\pm$0.05&    0.19$\pm$0.05&    0.14$\pm$0.04&    	         --&	0.37$\pm$0.04&    0.15$\pm$0.04&			--&		     --&    0.14$\pm$0.04\\
4733.93  &  [\ion{Fe}{iii}]  &         3F  &    0.12$\pm$0.04&    0.10$\pm$0.03&    0.06$\pm$0.03&    	         --&	0.21$\pm$0.05&    0.06$\pm$0.02&			--&		     --&    0.06$\pm$0.02\\
4754.83  &  [\ion{Fe}{iii}]  &         3F  &    0.07$\pm$0.03&    0.11$\pm$0.03&    0.12$\pm$0.04&    	         --&	0.12$\pm$0.04&    0.10$\pm$0.04&			--&		     --&    0.11$\pm$0.03\\
4861.33  &       \ion{H}{i}  &   H$\beta$  &    100$\pm$7  &    100$\pm$5  &    100$\pm$5  &    100$\pm$7  &    100$\pm$5  &    100$\pm$5  &    100$\pm$7  &    100$\pm$5  &    100$\pm$5  \\
4881.00  &  [\ion{Fe}{iii}]  &         2F  &    	     --&    0.17$\pm$0.04&    0.22$\pm$0.03&    	         --&    0.28$\pm$0.04&    0.23$\pm$0.04&    	            --&    	         --&    0.21$\pm$0.03\\
4958.91  &   [\ion{O}{iii}]  &         1F  &    80.9$\pm$5.9 &    108$\pm$5  &    116$\pm$5  &    75.0$\pm$5.5 &    103$\pm$5  &    114$\pm$5  &    124$\pm$9  &    117$\pm$5  &    115$\pm$5  \\
5006.84  &   [\ion{O}{iii}]  &         1F  &    233$\pm$17 &    318$\pm$14 &    348$\pm$16 &    228$\pm$17&    311$\pm$15 &    342$\pm$15 &    398$\pm$29 &    355$\pm$16 &    347$\pm$16 \\
5754.64  &    [\ion{N}{ii}]  &         3F  &    5.91$\pm$0.65&    1.67$\pm$0.13&    0.50$\pm$0.06&    3.30$\pm$0.34&    1.35$\pm$0.10&    0.71$\pm$0.05&    2.75$\pm$0.30&    1.06$\pm$0.08&    0.60$\pm$0.07\\
5875.64  &      \ion{He}{i}  &         11  &    14.5$\pm$1.4 &    13.6$\pm$0.8 &    13.4$\pm$0.8 &    11.4$\pm$1.0 &    12.4$\pm$0.7 &    13.3$\pm$0.8 &    15.1$\pm$1.4 &    13.6$\pm$0.8 &    13.1$\pm$0.7 \\
6300.30  &     [\ion{O}{i}]$^a$&       1F  &    4.85$\pm$0.45&    1.55$\pm$0.11&    0.57$\pm$0.06&    1.64$\pm$0.15&    0.96$\pm$0.07&    0.83$\pm$0.06&    3.89$\pm$0.36&    1.33$\pm$0.09&    0.71$\pm$0.05\\
6312.10  &   [\ion{S}{iii}]  &         3F  &    3.88$\pm$0.36&    2.08$\pm$0.15&	1.51$\pm$0.11&    2.03$\pm$0.18&	1.62$\pm$0.11&    1.53$\pm$0.11&	2.64$\pm$0.24&    1.74$\pm$0.12&	1.55$\pm$0.11\\
6548.03  &    [\ion{N}{ii}]  &         1F  &    11.3$\pm$0.9 &    10.6$\pm$0.5 &	10.4$\pm$0.5 &    12.9$\pm$1.0 &	13.2$\pm$0.7 &    13.9$\pm$0.7 &	7.19$\pm$0.64&    11.8$\pm$0.6 &	13.3$\pm$0.7 \\
6562.82  &       \ion{H}{i}  &  H$\alpha$  &    286$\pm$22 &    286$\pm$14 &    286$\pm$14 &    286$\pm$22 &    286$\pm$14 &    286$\pm$14 &    286$\pm$22 &    286$\pm$14 &    286$\pm$14 \\
6583.41  &    [\ion{N}{ii}]  &         1F  &    33.8$\pm$2.5 &    32.0$\pm$1.6 &    31.4$\pm$1.5&    38.3$\pm$2.9 &    40.0$\pm$2.0 &    41.6$\pm$2.0 &    20.9$\pm$1.6 &    36.0$\pm$1.8 &    40.5$\pm$2.0 \\
6678.15  &      \ion{He}{i}  &         46  &    3.51$\pm$0.32&    3.64$\pm$0.25&	3.59$\pm$0.25&    3.36$\pm$0.30&	3.50$\pm$0.24&    3.63$\pm$0.25&	3.38$\pm$0.30&    3.49$\pm$0.24&	3.57$\pm$0.25\\
6716.47  &    [\ion{S}{ii}]  &         2F  &    	     --&    1.64$\pm$0.12&	2.07$\pm$0.15&    1.52$\pm$0.14&	2.10$\pm$0.15&    2.28$\pm$0.16&	    	     --&    1.79$\pm$0.13&	2.33$\pm$0.16\\
6730.85  &    [\ion{S}{ii}]  &         2F  &    0.75$\pm$0.07&    2.68$\pm$0.19&	3.25$\pm$0.23&    2.21$\pm$0.20&	3.56$\pm$0.25&    4.02$\pm$0.28&	1.00$\pm$0.09&    3.38$\pm$0.24&	4.06$\pm$0.28\\
7135.78  &  [\ion{Ar}{iii}]  &         1F  &    17.0$\pm$1.5 &    14.1$\pm$0.8 &    13.3$\pm$0.72&    17.3$\pm$1.4 &    14.7$\pm$0.8 &    13.7$\pm$0.7 &    18.8$\pm$1.6 &    14.8$\pm$0.8 &    13.8$\pm$0.8 \\

%% file: tablaproplyd.tex
  \nel (\cmc)&                                 [\ion{S}{ii}] &    --   &          $>$14200 &              7600: &          $>$14200 &      7100: &          $>$14200 &     7030: & 2770$\pm$1300 & 2480$\pm$1050 \\ 
                  &        [\ion{Fe}{iii}]  ($\times10^3$)& 349.4$\pm$142.8 & 287.9$\pm$55.2 & 232.8$\pm$32.9 & 289.3$\pm$54.8 & 233.5$\pm$32.5 & 289.3$\pm$54.8 & 233.5$\pm$32.6 &  90.9$\pm$43.8 & 4.2$\pm$1.2 \\                       
                  & diag. \nel--\te$^b$ ($\times10^3$)&   398$\pm$121     & 289.1$\pm$89.9 & 244.4$\pm$68.4 & 284.4$\pm$79.6 & 239.1$\pm$66.9 & 283.3$\pm$79.3 & 237.2$\pm$63.4 &  93.3$\pm$12.9 &                    -- \\                   
                  &       adopted$^c$  ($\times10^3$)& 377.8$\pm$92.3   & 288.2$\pm$47.0 & 235.0$\pm$29.6 & 287.7$\pm$45.1 & 234.6$\pm$29.2 & 287.4$\pm$45.1 & 234.3$\pm$29.0 &  93.1$\pm$12.4 & 3.25$\pm$0.79 \\ 
         \te (K)&           [\ion{N}{ii}] & 8440$\pm$1060& 8410$\pm$760&   8550$\pm$680& 8340$\pm$810& 8430$\pm$640&   8310$\pm$730& 8400$\pm$660&  8570$\pm$530 & 10100$\pm$540\\                 
                  &          [\ion{O}{iii}] & 8300$\pm$400  & 8360$\pm$330&   8430$\pm$300& 8310$\pm$320& 8360$\pm$300&   8300$\pm$320& 8350$\pm$300&  8580$\pm$210 & 8780$\pm$190 \\
                  & diag. \nel--\te$^b$ & 8280$\pm$580  & 8390$\pm$520&   8430$\pm$490& 8360$\pm$560& 8370$\pm$510&   8360$\pm$440& 8360$\pm$480&  8540$\pm$260 &                      -- \\                      
															                    \multicolumn{11}{c}{}\\
                                                                                                             \multicolumn{11}{c}{Ionic abundances and \adfo}\\                                                                                                        
\multicolumn{2}{c}{\ioni{He}{+}$^d$} &10.96$\pm$0.02&10.96$\pm$0.02&10.97$\pm$0.02 &10.96$\pm$0.02&10.96$\pm$0.02&10.95$\pm$0.02 &10.96$\pm$0.02 &10.95$\pm$0.02 &10.96$\pm$0.02  \\
\multicolumn{2}{c}{\ioni{C}{2+}$^d$} & 8.25$\pm$0.12 & 8.27$\pm$0.12 & 8.29$\pm$0.12 & 8.27$\pm$0.12 & 8.29$\pm$0.12 & 8.27$\pm$0.12 & 8.29$\pm$0.12 & 8.33$\pm$0.11 & 8.35$\pm$0.07  \\
        \multicolumn{2}{c}{\ioni{N}{+}} & 7.83$\pm$0.14 & 7.72$\pm$0.09 & 7.63$\pm$0.08 & 7.74$\pm$0.10 & 7.65$\pm$0.08 & 7.75$\pm$0.09 & 7.65$\pm$0.08 & 7.34$\pm$0.06 & 6.78$\pm$0.05  \\
        \multicolumn{2}{c}{\ioni{O}{+}} &           --          &          --          &          --          &          --          &          --          &          --          &          --          &          --          & 7.64$\pm$0.11  \\
      \multicolumn{2}{c}{\ioni{O}{2+}} & 8.42$\pm$0.07 & 8.44$\pm$0.05 & 8.41$\pm$0.05 & 8.45$\pm$0.05 & 8.43$\pm$0.05 & 8.45$\pm$0.05 & 8.43$\pm$0.05 & 8.36$\pm$0.03 & 8.30$\pm$0.03  \\      
\multicolumn{2}{c}{\ioni{O}{2+}$^d$} & 8.37$\pm$0.14 & 8.39$\pm$0.11 & 8.39$\pm$0.11 & 8.38$\pm$0.11 & 8.39$\pm$0.11 & 8.38$\pm$0.11 & 8.39$\pm$0.11 & 8.40$\pm$0.11 & 8.46$\pm$0.11  \\
    \multicolumn{2}{c}{\ioni{Ne}{2+}} & 7.44$\pm$0.12 & 7.47$\pm$0.10 & 7.50$\pm$0.09 & 7.48$\pm$0.10 & 7.51$\pm$0.09 & 7.48$\pm$0.10 & 7.51$\pm$0.09 & 7.60$\pm$0.06 & 7.63$\pm$0.05  \\
        \multicolumn{2}{c}{\ioni{S}{+}} & 6.33$\pm$0.15 & 6.23$\pm$0.13 & 6.13$\pm$0.11 & 6.24$\pm$0.14 & 6.15$\pm$0.11 & 6.25$\pm$0.12 & 6.16$\pm$0.11 & 5.86$\pm$0.09 & 5.36$\pm$0.08  \\
      \multicolumn{2}{c}{\ioni{S}{2+}} & 7.15$\pm$0.11 & 7.11$\pm$0.09 & 7.06$\pm$0.09 & 7.11$\pm$0.09 & 7.07$\pm$0.09 & 7.11$\pm$0.09 & 7.07$\pm$0.09  & 6.89$\pm$0.06 & 6.78$\pm$0.06  \\
     \multicolumn{2}{c}{\ioni{Ar}{2+}} & 6.41$\pm$0.07 & 6.38$\pm$0.06 & 6.35$\pm$0.06 & 6.39$\pm$0.06 & 6.36$\pm$0.06 & 6.39$\pm$0.06 & 6.37$\pm$0.06  & 6.28$\pm$0.04 & 6.23$\pm$0.03  \\
     \multicolumn{2}{c}{\ioni{Fe}{2+}} & 5.48$\pm$0.14 & 5.56$\pm$0.15 & 5.54$\pm$0.11 & 5.57$\pm$0.15 & 5.56$\pm$0.12 & 5.57$\pm$0.15 & 5.56$\pm$0.12 & 5.53$\pm$0.10 & 5.34$\pm$0.11  \\
     \multicolumn{2}{c}{\adfo} & $-$0.06$\pm$0.15 & $-$0.05$\pm$0.13 & $-$0.02$\pm$0.12 & $-$0.06$\pm$0.13 & $-$0.04$\pm$0.13 & $-$0.07$\pm$0.13 & $-$0.04$\pm$0.12 & 0.04$\pm$0.11 & 0.16$\pm$0.11\\

%% file: tablaoprop.tex
  \nel (\cmc)&    [\ion{S}{ii}]        &     1850$\pm$950  &    3180$\pm$1600   & 4080$\pm$2300 &              --              &      6000: & 3760$\pm$2050\\ 
                  & [\ion{Fe}{iii}]        &                --            & 125900$\pm$39800 & 5950$\pm$1280 &       --              &              --           & 5380: \\ 
                  & diag. \nel-\te$^a$ & 128800$\pm$29900 &   53700$\pm$9950  &           --            & 158500$\pm$36800 & 24000$\pm$6100 &            --          \\ 
                  &    adopted$^b$    & 128800$\pm$29900 &   57900$\pm$9600  & 5650$\pm$1170 & 158500$\pm$36800 & 24000$\pm$6100 & 3760$\pm$2050\\ 

         \te (K)&           [\ion{N}{ii}] & 9470$\pm$1170&  8360$\pm$470& 10100$\pm$340&  10740$\pm$1660& 10100$\pm$770&  9590$\pm$580\\                 
                  &         [\ion{O}{iii}] &  9490$\pm$400 &   8600$\pm$200&   8870$\pm$190& 10870$\pm$560  & 10200$\pm$270&   8790$\pm$190\\
                  & diag. \nel-\te$^a$ & 9430$\pm$490 &   8560$\pm$240&          --            &  10770$\pm$690  & 10060$\pm$240&            --           \\                        

%% file: ms.bbl
\begin{thebibliography}{}

\bibitem[\protect\citeauthoryear{{Baldwin}, {Ferland}, {Martin}, {Corbin},
  {Cota}, {Peterson} \& {Slettebak}}{{Baldwin} et~al.}{1991}]{baldwinetal91}
{Baldwin} J.~A.,  {Ferland} G.~J.,  {Martin} P.~G.,  {Corbin} M.~R.,  {Cota}
  S.~A.,  {Peterson} B.~M.,    {Slettebak} A.,  1991, ApJ, 374, 580

\bibitem[\protect\citeauthoryear{{Bally}, {O'Dell} \& {McCaughrean}}{{Bally}
  et~al.}{2000}]{ballyetal00}
{Bally} J.,  {O'Dell} C.~R.,    {McCaughrean} M.~J.,  2000, AJ, 119, 2919

\bibitem[\protect\citeauthoryear{{Bally}, {Sutherland}, {Devine} \&
  {Johnstone}}{{Bally} et~al.}{1998}]{ballyetal98}
{Bally} J.,  {Sutherland} R.~S.,  {Devine} D.,    {Johnstone} D.,  1998, AJ,
  116, 293

\bibitem[\protect\citeauthoryear{{Beckwith}, {Sargent}, {Chini} \&
  {Guesten}}{{Beckwith} et~al.}{1990}]{beckwithetal90}
{Beckwith} S.~V.~W.,  {Sargent} A.~I.,  {Chini} R.~S.,    {Guesten} R.,  1990,
  $aj$, 99, 924

\bibitem[\protect\citeauthoryear{{Benjamin}, {Skillman} \& {Smits}}{{Benjamin}
  et~al.}{2002}]{benjaminetal02}
{Benjamin} R.~A.,  {Skillman} E.~D.,    {Smits} D.~P.,  2002, ApJ, 569, 288

\bibitem[\protect\citeauthoryear{{Blagrave}, {Martin}, {Rubin}, {Dufour},
  {Baldwin}, {Hester} \& {Walter}}{{Blagrave} et~al.}{2007}]{blagraveetal07}
{Blagrave} K.~P.~M.,  {Martin} P.~G.,  {Rubin} R.~H.,  {Dufour} R.~J.,
  {Baldwin} J.~A.,  {Hester} J.~J.,    {Walter} D.~K.,  2007, ApJ, 655, 299

\bibitem[\protect\citeauthoryear{{Brandner}, {Grebel}, {Chu}, {Dottori},
  {Brandl}, {Richling}, {Yorke}, {Points} \& {Zinnecker}}{{Brandner}
  et~al.}{2000}]{brandneretal00}
{Brandner} W.,  {Grebel} E.~K.,  {Chu} Y.,  {Dottori} H.,  {Brandl} B.,
  {Richling} S.,  {Yorke} H.~W.,  {Points} S.~D.,    {Zinnecker} H.,  2000, AJ,
  119, 292

\bibitem[\protect\citeauthoryear{{Cardelli}, {Clayton} \& {Mathis}}{{Cardelli}
  et~al.}{1989}]{cardellietal89}
{Cardelli} J.~A.,  {Clayton} G.~C.,    {Mathis} J.~S.,  1989, ApJ, 345, 245

\bibitem[\protect\citeauthoryear{{Chen}, {Bally}, {O'dell}, {McCaughrean},
  {Thompson}, {Rieke}, {Schneider} \& {Young}}{{Chen}
  et~al.}{1998}]{chenetal98}
{Chen} H.,  {Bally} J.,  {O'dell} C.~R.,  {McCaughrean} M.~J.,  {Thompson}
  R.~L.,  {Rieke} M.,  {Schneider} G.,    {Young} E.~T.,  1998, ApJ, 492, L173

\bibitem[\protect\citeauthoryear{{Churchwell}, {Wood}, {Felli} \&
  {Massi}}{{Churchwell} et~al.}{1987}]{churchwelletal87}
{Churchwell} E.,  {Wood} D.~O.~S.,  {Felli} M.,    {Massi} M.,  1987, ApJ, 321,
  516

\bibitem[\protect\citeauthoryear{{Costero} \& {Peimbert}}{{Costero} \&
  {Peimbert}}{1970}]{costeropeimbert70}
{Costero} R.,  {Peimbert} M.,  1970, Boletin de los Observatorios Tonantzintla
  y Tacubaya, 5, 229

\bibitem[\protect\citeauthoryear{{Davey}, {Storey} \& {Kisielius}}{{Davey}
  et~al.}{2000}]{daveyetal00}
{Davey} A.~R.,  {Storey} P.~J.,    {Kisielius} R.,  2000, A\&AS, 142, 85

\bibitem[\protect\citeauthoryear{{De Marco}, {O'Dell}, {Gelfond}, {Rubin} \&
  {Glover}}{{De Marco} et~al.}{2006}]{demarcoetal06}
{De Marco} O.,  {O'Dell} C.~R.,  {Gelfond} P.,  {Rubin} R.~H.,    {Glover}
  S.~C.~O.,  2006, AJ, 131, 2580

\bibitem[\protect\citeauthoryear{{Eisner} \& {Carpenter}}{{Eisner} \&
  {Carpenter}}{2006}]{eisnercarpenter06}
{Eisner} J.~A.,  {Carpenter} J.~M.,  2006, ApJ, 641, 1162

\bibitem[\protect\citeauthoryear{{Eisner}, {Plambeck}, {Carpenter}, {Corder},
  {Qi} \& {Wilner}}{{Eisner} et~al.}{2008}]{eisneretal08}
{Eisner} J.~A.,  {Plambeck} R.~L.,  {Carpenter} J.~M.,  {Corder} S.~A.,  {Qi}
  C.,    {Wilner} D.,  2008, ApJ, 683, 304

\bibitem[\protect\citeauthoryear{{Esteban}}{{Esteban}}{2002}]{esteban02}
{Esteban} C.,  2002, in {Henney} W.~J.,  {Franco} J.,   {Martos} M.,  eds,
  Revista Mexicana de Astronomia y Astrofisica Conference Series Vol.~12 of
  Revista Mexicana de Astronomia y Astrofisica Conference Series, {Are
  Temperature Fluctuations Out There?}.
pp 56--61

\bibitem[\protect\citeauthoryear{{Esteban}, {Peimbert}, {Garc{\'{\i}}a-Rojas},
  {Ruiz}, {Peimbert} \& {Rodr{\'{\i}}guez}}{{Esteban}
  et~al.}{2004}]{estebanetal04}
{Esteban} C.,  {Peimbert} M.,  {Garc{\'{\i}}a-Rojas} J.,  {Ruiz} M.~T.,
  {Peimbert} A.,    {Rodr{\'{\i}}guez} M.,  2004, MNRAS, 355, 229

\bibitem[\protect\citeauthoryear{{Esteban}, {Peimbert}, {Torres-Peimbert} \&
  {Escalante}}{{Esteban} et~al.}{1998}]{estebanetal98}
{Esteban} C.,  {Peimbert} M.,  {Torres-Peimbert} S.,    {Escalante} V.,  1998,
  MNRAS, 295, 401

\bibitem[\protect\citeauthoryear{{Ferland}, {Korista}, {Verner}, {Ferguson},
  {Kingdon} \& {Verner}}{{Ferland} et~al.}{1998}]{ferlandetal98}
{Ferland} G.~J.,  {Korista} K.~T.,  {Verner} D.~A.,  {Ferguson} J.~W.,
  {Kingdon} J.~B.,    {Verner} E.~M.,  1998, PASP, 110, 761

\bibitem[\protect\citeauthoryear{{Garc{\'{\i}}a-Arredondo}, {Henney} \&
  {Arthur}}{{Garc{\'{\i}}a-Arredondo} et~al.}{2001}]{garciaarredondoetal01}
{Garc{\'{\i}}a-Arredondo} F.,  {Henney} W.~J.,    {Arthur} S.~J.,  2001, ApJ,
  561, 830

\bibitem[\protect\citeauthoryear{{Garc{\'{\i}}a-D{\'{\i}}az} \&
  {Henney}}{{Garc{\'{\i}}a-D{\'{\i}}az} \& {Henney}}{2007}]{garciadiazhenney07}
{Garc{\'{\i}}a-D{\'{\i}}az} M.~T.,  {Henney} W.~J.,  2007, AJ, 133, 952

\bibitem[\protect\citeauthoryear{{Garc{\'{\i}}a-D{\'{\i}}az}, {Henney},
  {L{\'o}pez} \& {Doi}}{{Garc{\'{\i}}a-D{\'{\i}}az}
  et~al.}{2008}]{garciadiazetal08}
{Garc{\'{\i}}a-D{\'{\i}}az} M.~T.,  {Henney} W.~J.,  {L{\'o}pez} J.~A.,
  {Doi} T.,  2008, Rev. Mexicana Astron. Astrofis., 44, 181

\bibitem[\protect\citeauthoryear{{Garc{\'{\i}}a-Rojas} \&
  {Esteban}}{{Garc{\'{\i}}a-Rojas} \& {Esteban}}{2007}]{garciarojasesteban07}
{Garc{\'{\i}}a-Rojas} J.,  {Esteban} C.,  2007, ApJ, 670, 457

\bibitem[\protect\citeauthoryear{{Garc{\'{\i}}a-Rojas}, {Pe{\~n}a} \&
  {Peimbert}}{{Garc{\'{\i}}a-Rojas} et~al.}{2009}]{garciarojasetal09}
{Garc{\'{\i}}a-Rojas} J.,  {Pe{\~n}a} M.,    {Peimbert} A.,  2009, A\&A, 496,
  139

\bibitem[\protect\citeauthoryear{{Graham}, {Meaburn}, {Garrington}, {O'Brien},
  {Henney} \& {O'Dell}}{{Graham} et~al.}{2002}]{grahametal02}
{Graham} M.~F.,  {Meaburn} J.,  {Garrington} S.~T.,  {O'Brien} T.~J.,  {Henney}
  W.~J.,    {O'Dell} C.~R.,  2002, ApJ, 570, 222

\bibitem[\protect\citeauthoryear{{Harrington} \& {Feibelman}}{{Harrington} \&
  {Feibelman}}{1984}]{harringtonfeibelman84}
{Harrington} J.~P.,  {Feibelman} W.~A.,  1984, ApJ, 277, 716

\bibitem[\protect\citeauthoryear{{Henney}}{{Henney}}{2000}]{henney00}
{Henney} W.~J.,  2000, in {S.~J.~Arthur, N.~S.~Brickhouse, \& J.~Franco} ed.,
  Revista Mexicana de Astronomia y Astrofisica Conference Series Vol.~9 of
  Revista Mexicana de Astronomia y Astrofisica Conference Series, {Kinematics
  and Ionization of the Proplyd M42 177-341}.
pp 198--200

\bibitem[\protect\citeauthoryear{{Henney} \& {Arthur}}{{Henney} \&
  {Arthur}}{1998}]{henneyarthur98}
{Henney} W.~J.,  {Arthur} S.~J.,  1998, AJ, 116, 322

\bibitem[\protect\citeauthoryear{{Henney}, {Arthur}, {Williams} \&
  {Ferland}}{{Henney} et~al.}{2005}]{henneyetal05}
{Henney} W.~J.,  {Arthur} S.~J.,  {Williams} R.~J.~R.,    {Ferland} G.~J.,
  2005, ApJ, 621, 328

\bibitem[\protect\citeauthoryear{{Henney}, {Meaburn}, {Raga} \&
  {Massey}}{{Henney} et~al.}{1997}]{henneyetal97}
{Henney} W.~J.,  {Meaburn} J.,  {Raga} A.~C.,    {Massey} R.,  1997, A\&A, 324,
  656

\bibitem[\protect\citeauthoryear{{Henney} \& {O'Dell}}{{Henney} \&
  {O'Dell}}{1999}]{henneyodell99}
{Henney} W.~J.,  {O'Dell} C.~R.,  1999, AJ, 118, 2350

\bibitem[\protect\citeauthoryear{{Henney}, {O'Dell}, {Meaburn}, {Garrington} \&
  {Lopez}}{{Henney} et~al.}{2002}]{henneyetal02}
{Henney} W.~J.,  {O'Dell} C.~R.,  {Meaburn} J.,  {Garrington} S.~T.,    {Lopez}
  J.~A.,  2002, ApJ, 566, 315

\bibitem[\protect\citeauthoryear{{Hohenkerk} \& {Sinclair}}{{Hohenkerk} \&
  {Sinclair}}{1985}]{hohenkerksinclair85}
{Hohenkerk} C.~Y.,  {Sinclair} A.~T.,  1985, {The Computation of Angular
  Atmospheric Refraction at Large Zenith Angles}.
Royal Greenwich Observatory

\bibitem[\protect\citeauthoryear{{Johansson}, {Zethson}, {Hartman}, {Ekberg},
  {Ishibashi}, {Davidson} \& {Gull}}{{Johansson}
  et~al.}{2000}]{johanssonetal00}
{Johansson} S.,  {Zethson} T.,  {Hartman} H.,  {Ekberg} J.~O.,  {Ishibashi} K.,
   {Davidson} K.,    {Gull} T.,  2000, A\&A, 361, 977

\bibitem[\protect\citeauthoryear{{Johnstone}, {Hollenbach} \&
  {Bally}}{{Johnstone} et~al.}{1998}]{johnstoneetal98}
{Johnstone} D.,  {Hollenbach} D.,    {Bally} J.,  1998, ApJ, 499, 758

\bibitem[\protect\citeauthoryear{{Kingdon} \& {Ferland}}{{Kingdon} \&
  {Ferland}}{1995}]{kingdonferland95}
{Kingdon} J.,  {Ferland} G.~J.,  1995, ApJ, 442, 714

\bibitem[\protect\citeauthoryear{{Lada}, {Muench}, {Haisch} Jr., {Lada},
  {Alves}, {Tollestrup} \& {Willner}}{{Lada} et~al.}{2000}]{ladaetal00}
{Lada} C.~J.,  {Muench} A.~A.,  {Haisch} Jr. K.~E.,  {Lada} E.~A.,  {Alves}
  J.~F.,  {Tollestrup} E.~V.,    {Willner} S.~P.,  2000, AJ, 120, 3162

\bibitem[\protect\citeauthoryear{{Lanz} \& {Hubeny}}{{Lanz} \&
  {Hubeny}}{2003}]{lanzhubeny03}
{Lanz} T.,  {Hubeny} I.,  2003, ApJS, 146, 417

\bibitem[\protect\citeauthoryear{{Laques} \& {Vidal}}{{Laques} \&
  {Vidal}}{1979}]{laquesvidal79}
{Laques} P.,  {Vidal} J.~L.,  1979, A\&A, 73, 97

\bibitem[\protect\citeauthoryear{{Liu}}{{Liu}}{2002}]{liu02}
{Liu} X.-W.,  2002, in {Henney} W.~J.,  {Franco} J.,   {Martos} M.,  eds,
  Revista Mexicana de Astronomia y Astrofisica Conference Series Vol.~12 of
  Revista Mexicana de Astronomia y Astrofisica Conference Series, {Optical
  Recombination Lines and Temperature Fluctuations}.
pp 70--76

\bibitem[\protect\citeauthoryear{{Liu}}{{Liu}}{2006}]{liu06}
{Liu} X.-W.,  2006, in {Barlow} M.~J.,  {M{\'e}ndez} R.~H.,  eds, Planetary
  Nebulae in our Galaxy and Beyond Vol.~234 of IAU Symposium, Optical
  recombination lines as probes of conditions in planetary nebulae.
pp 219--226

\bibitem[\protect\citeauthoryear{{Liu}, {Storey}, {Barlow}, {Danziger}, {Cohen}
  \& {Bryce}}{{Liu} et~al.}{2000}]{liuetal00}
{Liu} X.-W.,  {Storey} P.~J.,  {Barlow} M.~J.,  {Danziger} I.~J.,  {Cohen} M.,
    {Bryce} M.,  2000, MNRAS, 312, 585

\bibitem[\protect\citeauthoryear{{Luridiana}, {Morriset} \& {Shaw}}{{Luridiana}
  et~al.}{2011}]{luridianaetal11}
{Luridiana} V.,  {Morriset} C.,    {Shaw} R.~A.,  2011, in {A.~Manchado \&
  L.~Stanghellini} ed., Planetary Nebulae: An Eye to the Future Vol.~283 of IAU
  Symposium, Pyneb: a new software for the analysis of emission lines.
Cambridge University Press

\bibitem[\protect\citeauthoryear{{Massey} \& {Meaburn}}{{Massey} \&
  {Meaburn}}{1993}]{masseymeaburn93}
{Massey} R.~M.,  {Meaburn} J.,  1993, MNRAS, 262, L48

\bibitem[\protect\citeauthoryear{{Massey} \& {Meaburn}}{{Massey} \&
  {Meaburn}}{1995}]{masseymeaburn95}
{Massey} R.~M.,  {Meaburn} J.,  1995, MNRAS, 273, 615

\bibitem[\protect\citeauthoryear{{McCaughrean} \& {O'Dell}}{{McCaughrean} \&
  {O'Dell}}{1996}]{mccaughreanodell96}
{McCaughrean} M.~J.,  {O'Dell} C.~R.,  1996, AJ, 111, 1977

\bibitem[\protect\citeauthoryear{{McCullough}, {Fugate}, {Christou},
  {Ellerbroek}, {Higgins}, {Spinhirne}, {Cleis} \& {Moroney}}{{McCullough}
  et~al.}{1995}]{mcculloughetal95}
{McCullough} P.~R.,  {Fugate} R.~Q.,  {Christou} J.~C.,  {Ellerbroek} B.~L.,
  {Higgins} C.~H.,  {Spinhirne} J.~M.,  {Cleis} R.~A.,    {Moroney} J.~F.,
  1995, ApJ, 438, 394

\bibitem[\protect\citeauthoryear{{Meaburn}}{{Meaburn}}{1988}]{meaburn88}
{Meaburn} J.,  1988, MNRAS, 233, 791

\bibitem[\protect\citeauthoryear{{Meaburn}, {Massey}, {Raga} \&
  {Clayton}}{{Meaburn} et~al.}{1993}]{meaburnetal93}
{Meaburn} J.,  {Massey} R.~M.,  {Raga} A.~C.,    {Clayton} C.~A.,  1993, MNRAS,
  260, 625

\bibitem[\protect\citeauthoryear{{Mesa-Delgado}, {Esteban} \&
  {Garc{\'{\i}}a-Rojas}}{{Mesa-Delgado} et~al.}{2008}]{mesadelgadoetal08}
{Mesa-Delgado} A.,  {Esteban} C.,    {Garc{\'{\i}}a-Rojas} J.,  2008, ApJ, 675,
  389

\bibitem[\protect\citeauthoryear{{Mesa-Delgado}, {L{\'o}pez-Mart{\'{\i}}n},
  {Esteban}, {Garc{\'{\i}}a-Rojas} \& {Luridiana}}{{Mesa-Delgado}
  et~al.}{2009}]{mesadelgadoetal09a}
{Mesa-Delgado} A.,  {L{\'o}pez-Mart{\'{\i}}n} L.,  {Esteban} C.,
  {Garc{\'{\i}}a-Rojas} J.,    {Luridiana} V.,  2009, MNRAS, 394, 693

\bibitem[\protect\citeauthoryear{{Mesa-Delgado}, {N{\'u}{\~n}ez-D{\'{\i}}az},
  {Esteban}, {L{\'o}pez-Mart{\'{\i}}n} \& {Garc{\'{\i}}a-Rojas}}{{Mesa-Delgado}
  et~al.}{2011}]{mesadelgadoetal11}
{Mesa-Delgado} A.,  {N{\'u}{\~n}ez-D{\'{\i}}az} M.,  {Esteban} C.,
  {L{\'o}pez-Mart{\'{\i}}n} L.,    {Garc{\'{\i}}a-Rojas} J.,  2011, MNRAS, 417,
  420

\bibitem[\protect\citeauthoryear{{Neves}, {Santos}, {Sousa}, {Correia} \&
  {Israelian}}{{Neves} et~al.}{2009}]{nevesetal09}
{Neves} V.,  {Santos} N.~C.,  {Sousa} S.~G.,  {Correia} A.~C.~M.,
  {Israelian} G.,  2009, A\&A, 497, 563

\bibitem[\protect\citeauthoryear{{Nguyen}, {Viti} \& {Williams}}{{Nguyen}
  et~al.}{2002}]{nguyenetal02}
{Nguyen} T.~K.,  {Viti} S.,    {Williams} D.~A.,  2002, A\&A, 387, 1083

\bibitem[\protect\citeauthoryear{{O'Dell}}{{O'Dell}}{1998}]{odell98}
{O'Dell} C.~R.,  1998, AJ, 115, 263

\bibitem[\protect\citeauthoryear{{O'Dell} \& {Henney}}{{O'Dell} \&
  {Henney}}{2008}]{odellhenney08}
{O'Dell} C.~R.,  {Henney} W.~J.,  2008, AJ, 136, 1566

\bibitem[\protect\citeauthoryear{{O'Dell} \& {Wen}}{{O'Dell} \&
  {Wen}}{1994}]{odellwen94}
{O'Dell} C.~R.,  {Wen} Z.,  1994, ApJ, 436, 194

\bibitem[\protect\citeauthoryear{{O'Dell}, {Wen} \& {Hu}}{{O'Dell}
  et~al.}{1993}]{odelletal93}
{O'Dell} C.~R.,  {Wen} Z.,    {Hu} X.,  1993, ApJ, 410, 696

\bibitem[\protect\citeauthoryear{{O'Dell} \& {Wong}}{{O'Dell} \&
  {Wong}}{1996}]{odellwong96}
{O'Dell} C.~R.,  {Wong} K.,  1996, AJ, 111, 846

\bibitem[\protect\citeauthoryear{{O'Dell} \& {Yusef-Zadeh}}{{O'Dell} \&
  {Yusef-Zadeh}}{2000}]{odellyusefzadeh00}
{O'Dell} C.~R.,  {Yusef-Zadeh} F.,  2000, AJ, 120, 382

\bibitem[\protect\citeauthoryear{{Oke}}{{Oke}}{1990}]{oke90}
{Oke} J.~B.,  1990, AJ, 99, 1621

\bibitem[\protect\citeauthoryear{{Pauldrach}, {Hoffmann} \&
  {Lennon}}{{Pauldrach} et~al.}{2001}]{pauldrachetal01}
{Pauldrach} A.~W.~A.,  {Hoffmann} T.~L.,    {Lennon} M.,  2001, A\&A, 375, 161

\bibitem[\protect\citeauthoryear{{Peimbert} \& {Peimbert}}{{Peimbert} \&
  {Peimbert}}{2005}]{apeimbertpeimbert05}
{Peimbert} A.,  {Peimbert} M.,  2005, in {Torres-Peimbert} D.,  {MacAlpine} G.,
   eds, Rev. Mexicana Astron. Astrofis. Conf. Ser. Vol.~23, {Oxygen
  Recombination Line Abundances in Gaseous Nebulae}.
p.~9

\bibitem[\protect\citeauthoryear{{Peimbert}}{{Peimbert}}{1967}]{peimbert67}
{Peimbert} M.,  1967, ApJ, 150, 825

\bibitem[\protect\citeauthoryear{{Peimbert} \& {Peimbert}}{{Peimbert} \&
  {Peimbert}}{2006}]{peimbertapeimbert06}
{Peimbert} M.,  {Peimbert} A.,  2006, in {Barlow} M.~J.,  {M{\'e}ndez} R.~H.,
  eds, Planetary Nebulae in our Galaxy and Beyond Vol.~234 of IAU Symposium,
  {Temperature Variations and Chemical Abundances in Planetary Nebulae}.
pp 227--234

\bibitem[\protect\citeauthoryear{{Peimbert}, {Storey} \&
  {Torres-Peimbert}}{{Peimbert} et~al.}{1993}]{peimbertetal93}
{Peimbert} M.,  {Storey} P.~J.,    {Torres-Peimbert} S.,  1993, ApJ, 414, 626

\bibitem[\protect\citeauthoryear{{Porter}, {Bauman}, {Ferland} \&
  {MacAdam}}{{Porter} et~al.}{2005}]{porteretal05}
{Porter} R.~L.,  {Bauman} R.~P.,  {Ferland} G.~J.,    {MacAdam} K.~B.,  2005,
  ApJ, 622, L73

\bibitem[\protect\citeauthoryear{{Porter}, {Ferland} \& {MacAdam}}{{Porter}
  et~al.}{2007}]{porteretal07}
{Porter} R.~L.,  {Ferland} G.~J.,    {MacAdam} K.~B.,  2007, ApJ, 657, 327

\bibitem[\protect\citeauthoryear{{Quinet}}{{Quinet}}{1996}]{quinet96}
{Quinet} P.,  1996, A\&AS, 116, 573

\bibitem[\protect\citeauthoryear{{Ricci}, {Robberto} \& {Soderblom}}{{Ricci}
  et~al.}{2008}]{riccietal08}
{Ricci} L.,  {Robberto} M.,    {Soderblom} D.~R.,  2008, AJ, 136, 2136

\bibitem[\protect\citeauthoryear{{Richling} \& {Yorke}}{{Richling} \&
  {Yorke}}{1998}]{richlingyorke98}
{Richling} S.,  {Yorke} H.~W.,  1998, A\&A, 340, 508

\bibitem[\protect\citeauthoryear{{Robberto}, {Beckwith} \&
  {Panagia}}{{Robberto} et~al.}{2002}]{robbertoetal02}
{Robberto} M.,  {Beckwith} S.~V.~W.,    {Panagia} N.,  2002, ApJ, 578, 897

\bibitem[\protect\citeauthoryear{{Robberto}, {Beckwith}, {Panagia}, {Patel},
  {Herbst}, {Ligori}, {Custo}, {Boccacci} \& {Bertero}}{{Robberto}
  et~al.}{2005}]{robbertoetal05}
{Robberto} M.,  {Beckwith} S.~V.~W.,  {Panagia} N.,  {Patel} S.~G.,  {Herbst}
  T.~M.,  {Ligori} S.,  {Custo} A.,  {Boccacci} P.,    {Bertero} M.,  2005, AJ,
  129, 1534

\bibitem[\protect\citeauthoryear{{Rodr{\'{\i}}guez} \&
  {Garc{\'{\i}}a-Rojas}}{{Rodr{\'{\i}}guez} \&
  {Garc{\'{\i}}a-Rojas}}{2010}]{rodriguezgarciarojas10}
{Rodr{\'{\i}}guez} M.,  {Garc{\'{\i}}a-Rojas} J.,  2010, ApJ, 708, 1551

\bibitem[\protect\citeauthoryear{{Rost}, {Eckart} \& {Ott}}{{Rost}
  et~al.}{2008}]{rostetal08}
{Rost} S.,  {Eckart} A.,    {Ott} T.,  2008, A\&A, 485, 107

\bibitem[\protect\citeauthoryear{{Roth}, {Kelz}, {Fechner}, {Hahn}, {Bauer},
  {Becker}, {B{\"o}hm}, {Christensen}, {Dionies}, {Paschke}, {Popow}, {Wolter},
  {Schmoll}, {Laux} \& {Altmann}}{{Roth} et~al.}{2005}]{rothetal2005}
{Roth} M.~M.,  {Kelz} A.,  {Fechner} T.,  {Hahn} T.,  {Bauer} S.-M.,  {Becker}
  T.,  {B{\"o}hm} P.,  {Christensen} L.,  {Dionies} F.,  {Paschke} J.,  {Popow}
  E.,  {Wolter} D.,  {Schmoll} J.,  {Laux} U.,    {Altmann} W.,  2005, PASP,
  117, 620

\bibitem[\protect\citeauthoryear{{Rubin}}{{Rubin}}{1989}]{rubin89}
{Rubin} R.~H.,  1989, ApJS, 69, 897

\bibitem[\protect\citeauthoryear{{Rubin}, {Martin}, {Dufour}, {Ferland},
  {Blagrave}, {Liu}, {Nguyen} \& {Baldwin}}{{Rubin} et~al.}{2003}]{rubinetal03}
{Rubin} R.~H.,  {Martin} P.~G.,  {Dufour} R.~J.,  {Ferland} G.~J.,  {Blagrave}
  K.~P.~M.,  {Liu} X.-W.,  {Nguyen} J.~F.,    {Baldwin} J.~A.,  2003, MNRAS,
  340, 362

\bibitem[\protect\citeauthoryear{{Sawey} \& {Berrington}}{{Sawey} \&
  {Berrington}}{1993}]{saweyberrington93}
{Sawey} P.~M.~J.,  {Berrington} K.~A.,  1993, Atomic Data and Nuclear Data
  Tables, 55, 81

\bibitem[\protect\citeauthoryear{{Shaw} \& {Dufour}}{{Shaw} \&
  {Dufour}}{1995}]{shawdufour95}
{Shaw} R.~A.,  {Dufour} R.~J.,  1995, PASP, 107, 896

\bibitem[\protect\citeauthoryear{{Shuping}, {Kassis}, {Morris}, {Smith} \&
  {Bally}}{{Shuping} et~al.}{2006}]{shupingetal06}
{Shuping} R.~Y.,  {Kassis} M.,  {Morris} M.,  {Smith} N.,    {Bally} J.,  2006,
  ApJ, 644, L71

\bibitem[\protect\citeauthoryear{{Shuping}, {Patience}, {Bally}, {Morris},
  {Larkin} \& {Macintosh}}{{Shuping} et~al.}{2003}]{shupingetal03}
{Shuping} R.~Y.,  {Patience} J.,  {Bally} J.,  {Morris} M.,  {Larkin} J.~E.,
  {Macintosh} B.~A.,  2003, in {P.~Guhathakurta} ed., Society of Photo-Optical
  Instrumentation Engineers (SPIE) Conference Series Vol.~4834 of Society of
  Photo-Optical Instrumentation Engineers (SPIE) Conference Series, {Keck
  near-infrared observations of the Orion Proplyds: Initial results}.
pp 364--374

\bibitem[\protect\citeauthoryear{{Sim{\'o}n-D{\'{\i}}az}, {Herrero}, {Esteban}
  \& {Najarro}}{{Sim{\'o}n-D{\'{\i}}az} et~al.}{2006}]{simondiazetal06}
{Sim{\'o}n-D{\'{\i}}az} S.,  {Herrero} A.,  {Esteban} C.,    {Najarro} F.,
  2006, A\&A, 448, 351

\bibitem[\protect\citeauthoryear{{Sim{\'o}n-D{\'{\i}}az} \&
  {Stasi{\'n}ska}}{{Sim{\'o}n-D{\'{\i}}az} \&
  {Stasi{\'n}ska}}{2011}]{simondiazstasinska11}
{Sim{\'o}n-D{\'{\i}}az} S.,  {Stasi{\'n}ska} G.,  2011, A\&A, 526, A48

\bibitem[\protect\citeauthoryear{{Smith}, {Bally} \& {Morse}}{{Smith}
  et~al.}{2003}]{smithetal03}
{Smith} N.,  {Bally} J.,    {Morse} J.~A.,  2003, ApJ, 587, L105

\bibitem[\protect\citeauthoryear{{Smith}, {Bally}, {Shuping}, {Morris} \&
  {Kassis}}{{Smith} et~al.}{2005}]{smithetal05}
{Smith} N.,  {Bally} J.,  {Shuping} R.~Y.,  {Morris} M.,    {Kassis} M.,  2005,
  AJ, 130, 1763

\bibitem[\protect\citeauthoryear{{Stapelfeldt}, {Sahai}, {Werner} \&
  {Trauger}}{{Stapelfeldt} et~al.}{1997}]{stapelfeldtetal97}
{Stapelfeldt} K.,  {Sahai} R.,  {Werner} M.,    {Trauger} J.,  1997, in
  {D.~Soderblom} ed., Planets Beyond the Solar System and the Next Generation
  of Space Missions Vol.~119 of Astronomical Society of the Pacific Conference
  Series, {An HST Imaging Search for Circumstellar Matter in Young Nebulous
  Clusters}.
p.~131

\bibitem[\protect\citeauthoryear{{Stasi{\'n}ska}, {Tenorio-Tagle},
  {Rodr{\'{\i}}guez} \& {Henney}}{{Stasi{\'n}ska}
  et~al.}{2007}]{stasinskaetal07}
{Stasi{\'n}ska} G.,  {Tenorio-Tagle} G.,  {Rodr{\'{\i}}guez} M.,    {Henney}
  W.~J.,  2007, A\&A, 471, 193

\bibitem[\protect\citeauthoryear{{Stecklum}, {Henning}, {Feldt}, {Hayward},
  {Hoare}, {Hofner} \& {Richter}}{{Stecklum} et~al.}{1998}]{stecklumetal98}
{Stecklum} B.,  {Henning} T.,  {Feldt} M.,  {Hayward} T.~L.,  {Hoare} M.~G.,
  {Hofner} P.,    {Richter} S.,  1998, AJ, 115, 767

\bibitem[\protect\citeauthoryear{{Storey}}{{Storey}}{1994}]{storey94}
{Storey} P.~J.,  1994, A\&A, 282, 999

\bibitem[\protect\citeauthoryear{{Storey} \& {Hummer}}{{Storey} \&
  {Hummer}}{1995}]{storeyhummer95}
{Storey} P.~J.,  {Hummer} D.~G.,  1995, MNRAS, 272, 41

\bibitem[\protect\citeauthoryear{{Storzer} \& {Hollenbach}}{{Storzer} \&
  {Hollenbach}}{1998}]{storzerhollenbach98}
{Storzer} H.,  {Hollenbach} D.,  1998, ApJ, 502, L71

\bibitem[\protect\citeauthoryear{{St{\"o}rzer} \& {Hollenbach}}{{St{\"o}rzer}
  \& {Hollenbach}}{1999}]{storzerhollenbach99}
{St{\"o}rzer} H.,  {Hollenbach} D.,  1999, ApJ, 515, 669

\bibitem[\protect\citeauthoryear{{Sutherland}}{{Sutherland}}{1997}]{sutherland%
97}
{Sutherland} R.~S.,  1997, in {D.~T.~Wickramasinghe, G.~V.~Bicknell, \&
  L.~Ferrario} ed., IAU Colloq. 163: Accretion Phenomena and Related Outflows
  Vol.~121 of Astronomical Society of the Pacific Conference Series, {A Two
  Dimensional Photoionization Front Model for YSO Envelopes in the Orion
  Nebula}.
p.~566

\bibitem[\protect\citeauthoryear{{Tsamis}, {Barlow}, {Liu}, {Danziger} \&
  {Storey}}{{Tsamis} et~al.}{2003}]{tsamisetal03}
{Tsamis} Y.~G.,  {Barlow} M.~J.,  {Liu} X.-W.,  {Danziger} I.~J.,    {Storey}
  P.~J.,  2003, MNRAS, 338, 687

\bibitem[\protect\citeauthoryear{{Tsamis} \& {P{\'e}quignot}}{{Tsamis} \&
  {P{\'e}quignot}}{2005}]{tsamispequignot05}
{Tsamis} Y.~G.,  {P{\'e}quignot} D.,  2005, MNRAS, 364, 687

\bibitem[\protect\citeauthoryear{{Tsamis} \& {Walsh}}{{Tsamis} \&
  {Walsh}}{2011}]{tsamiswalsh11}
{Tsamis} Y.~G.,  {Walsh} J.~R.,  2011, MNRAS, 417, 2072

\bibitem[\protect\citeauthoryear{{Tsamis}, {Walsh}, {P{\'e}quignot}, {Barlow},
  {Danziger} \& {Liu}}{{Tsamis} et~al.}{2008}]{tsamisetal08}
{Tsamis} Y.~G.,  {Walsh} J.~R.,  {P{\'e}quignot} D.,  {Barlow} M.~J.,
  {Danziger} I.~J.,    {Liu} X.-W.,  2008, MNRAS, 386, 22

\bibitem[\protect\citeauthoryear{{Tsamis}, {Walsh}, {V{\'{\i}}lchez} \&
  {P{\'e}quignot}}{{Tsamis} et~al.}{2011}]{tsamisetal11}
{Tsamis} Y.~G.,  {Walsh} J.~R.,  {V{\'{\i}}lchez} J.~M.,    {P{\'e}quignot} D.,
   2011, MNRAS, 412, 1367

\bibitem[\protect\citeauthoryear{{Vasconcelos}, {Cerqueira}, {Plana}, {Raga} \&
  {Morisset}}{{Vasconcelos} et~al.}{2005}]{vasconcelosetal05}
{Vasconcelos} M.~J.,  {Cerqueira} A.~H.,  {Plana} H.,  {Raga} A.~C.,
  {Morisset} C.,  2005, AJ, 130, 1707

\bibitem[\protect\citeauthoryear{{Vasconcelos}, {Cerqueira} \&
  {Raga}}{{Vasconcelos} et~al.}{2011}]{vasconcelosetal11}
{Vasconcelos} M.~J.,  {Cerqueira} A.~H.,    {Raga} A.~C.,  2011, A\&A, 527, A86

\bibitem[\protect\citeauthoryear{{Viegas} \& {Clegg}}{{Viegas} \&
  {Clegg}}{1994}]{viegasclegg94}
{Viegas} S.~M.,  {Clegg} R.~E.~S.,  1994, MNRAS, 271, 993

\bibitem[\protect\citeauthoryear{{Walsh} \& {Roy}}{{Walsh} \&
  {Roy}}{1990}]{walshroy90}
{Walsh} J.~R.,  {Roy} J.~R.,  1990, in {D.~Baade \& P.~J.~Grosbol} ed.,
  European Southern Observatory Conference and Workshop Proceedings Vol.~34,
  {Area Spectroscopy and Correction for Differential Atmospheric Refraction}.
p.~95

\bibitem[\protect\citeauthoryear{{Williams} \& {Cieza}}{{Williams} \&
  {Cieza}}{2011}]{williamscieza11}
{Williams} J.~P.,  {Cieza} L.~A.,  2011, ARA\&A, 49, 67

\bibitem[\protect\citeauthoryear{{Wright}, {Drake}, {Drew}, {Guarcello},
  {Gutermuth}, {Hora} \& {Kraemer}}{{Wright} et~al.}{2012}]{wrightetal12}
{Wright} N.~J.,  {Drake} J.~J.,  {Drew} J.~E.,  {Guarcello} M.~G.,  {Gutermuth}
  R.~A.,  {Hora} J.~L.,    {Kraemer} K.~E.,  2012, $apjl$, 746, L21

\bibitem[\protect\citeauthoryear{{Yusef-Zadeh}, {Biretta} \&
  {Geballe}}{{Yusef-Zadeh} et~al.}{2005}]{yusefzadehetal05}
{Yusef-Zadeh} F.,  {Biretta} J.,    {Geballe} T.~R.,  2005, AJ, 130, 1171

\bibitem[\protect\citeauthoryear{{Zhang}}{{Zhang}}{1996}]{zhang96}
{Zhang} H.,  1996, A\&AS, 119, 523

\end{thebibliography}
